\documentclass[Journal,12pt,draftclsnofoot,onecolumn]{IEEEtran}
\usepackage{setspace}
\usepackage{cite}
\usepackage{graphicx,color,epsfig,rotating}
\usepackage{amsfonts,amsmath,amssymb}
\usepackage{algorithm,algorithmic}
\usepackage{subfigure}
\usepackage{url}
\usepackage{epsfig}
\usepackage{epstopdf}
\makeatletter
\def\ps@headings{%
\def\@oddhead{\mbox{}\scriptsize\rightmark \hfil \thepage}%
\def\@evenhead{\scriptsize\thepage \hfil \leftmark\mbox{}}%
\def\@oddfoot{}%
\def\@evenfoot{}}
\makeatother
\usepackage{url}

\def\ps@headings{%
\def\@oddhead{\mbox{}\scriptsize\rightmark \hfil \thepage}%
\def\@evenhead{\scriptsize\thepage \hfil \leftmark\mbox{}}%
\def\@oddfoot{}%
\def\@evenfoot{}}
\makeatother
\usepackage{color}

\begin{document}
\title{\LARGE{Base-Station Assisted Device-to-Device Communications for High-Throughput Wireless Video Networks}}
\author{Negin Golrezaei ~\IEEEmembership{Student Member}~IEEE,
Parisa Mansourifard ~\IEEEmembership{Student Member}~IEEE,
        Andreas F.~Molisch ~\IEEEmembership{Fellow}~IEEE,
        Alexandros G.~Dimakis,~\IEEEmembership{Member}~IEEE,
        \\ Deptartment of Electrical Engineering\\
University of Southern California\\
Los Angeles, CA, USA\\
emails:~\{golrezae,parisama,dimakis,molisch\}@usc.edu
        }
        \maketitle

\begin{abstract}
We propose a new scheme for increasing the throughput of video files in cellular communications systems. This scheme exploits (i) the redundancy of user requests as well as (ii) the considerable storage capacity of smartphones and tablets. Users cache popular video files and - after receiving requests from other users - serve these requests via device-to-device localized transmissions. The file placement is optimal when a central control knows a priori the locations of wireless devices when file requests occur. However, even a purely random caching scheme shows only a minor performance loss compared to such a "genie-aided" scheme. We then analyze the optimal collaboration distance, trading off frequency reuse with the probability of finding a requested file within the collaboration distance. We show that an improvement of spectral efficiency of one to two orders of magnitude is possible, even if there is not very high redundancy in video requests.

\end{abstract}

\section{Introduction}

Wireless video has been one of the main drivers of wireless (cellular) data traffic, and its importance is going to grow.
One of the main drivers of such traffic will be mobile video. Originally, wireless video mostly implied short video clips (YouTube or news channels) on the very small screens of smartphones. The recent popularity of tablets and large-screen phones has opened the possibility for watching feature-length movies at high resolution on mobile devices.
This increases the amount of data that has to be transmitted, and leads
to predictions that video will soon account for the majority of all wireless data traffic. According to recent industry estimates~\cite{cisco66} the traffic generated by video
delivery requests will quickly outpace mobile web content and lead to an increase in wireless data traffic by two orders of magnitude. These developments, while enabling new business models and possibly consumer satisfaction, threaten to completely clog up the already overburdened cellular networks.

Traditional methods for increasing cellular capacity are
\begin{itemize}
\item \emph{Improvement of the physical-layer link capacity between transmitter and receiver.}
However, fourth-generation cellular systems (LTE-Advanced) and wireless LANs (IEEE 802.11n/ac), use a physical layer (MIMO-OFDM with near-capacity-achieving codes) whose spectral efficiency is close to the theoretical limits, so that further improvements will be very limited.
\item \emph{Use of additional spectrum.}
In the best case we can increase the available spectrum by a factor of $2$, which is insufficient to satisfy the increased demand for data.
\item \emph{Decrease of the cell size to improve the area spectral efficiency.}
While promising, this approach suffers from the high costs of establishing new cell sites and providing the associated backhaul capacity \cite{surveyfemto}.
\end{itemize}

In this paper, we propose and analyze a novel architecture to improve the throughput of video transmission in cellular networks, based on (i) caching of popular video files in cellphones and (ii) base station controlled device-to-device (D2D) communications. Our architecture exploits the large storage available on modern smartphones to cache video files that might be requested by other users. Base stations keep track of the cache content and direct requests to the nearest smartphone that has the desired file, which is then transmitted via a D2D link. Since the distance between the requesting user and the user 
with the stored file will be small in most cases, multiple D2D links can be operated on the same time/frequency resources within one cell. This in turn leads to a dramatic increase in spectral efficiency.

\emph{Existing Literature:} Exploiting the redundancy of video requests in cellular networks is not new, but in the past has been used in fundamentally different ways compared to our approach. Wireless TV, such as MediaFlow, was exploiting the broadcast effect to supply the same video stream to many users simultaneously.
However, consumers expect \emph{on demand} capability for cellular transmission. Caching of popular files at the base stations, or at mobile switching centers is discussed, e.g., in \cite{ahlehagh2012hierarchical},
but it only helps to reduce the strain on the cellular backhaul, without alleviating the "on air" congestion. However, the efficiency of this approach is limited by the number of files that can be stored on a single device. A pioneering precoding scheme that works in conjunction with broadcast from the BS was recently introduced by \cite{maddah2013decentralized}. Caching files that are anticipated to be useful for a particular user on his/her personal device is another popular approach that was pioneered for set top boxes and has recently been applied to mobile devices \cite{chellouche2012home}. In contrast, we will present an approach that allows many devices to pool their caching resources, and exchange their stored files locally. This creates a much larger "virtual" cache, and thus dramatically increases the probability
of finding the request files in the virtual cache.
\footnote{An alternative, complementary, approach to dealing with the backhaul bottleneck was recently proposed in our recent prior work \cite{femtocaching}, \cite{golrezaei2012wireless}, where femto-base stations are replaced by small base stations with high storage capacity but low backhaul capacity.}

Also the use of D2D communications is not new. The literature on ad-hoc networks without central control is particularly rich, and we refrain from even attempting to give an overview of all the existing papers. Standardized commercial systems have recently been introduced (WiFi Direct \cite{wifi}), or have been proposed (FlashLinQ \cite{wu2010flashlinq}). The use of base-station controlled D2D has recently gathered the interest of the cellular industry, and has been reviewed, e.g., in \cite{doppler2010mode}. The central knowledge of the channel states and information requests has been shown to greatly improve the performance. However, these systems aim to establish direct communication between arbitrary users (e.g., friends that want to chat to each other), and thus cannot influence the distance that needs to be covered. Our proposed system allows the BS to influence which nodes communicate with each other, leading to a completely different mathematical formulation and performance results. In particular, we can show that the throughput in our system is not limited by the "Gupta-Kumar" limit \cite{gupta2000capacity}. A somewhat different approach was recently introduced under the name "microcast" by Keller et al. \cite{keller2012microcast}, where multiple devices download from the BS and then combine the gleaned information. The difference is that micro casting speeds up the download for a particular user, but does not necessarily increase the sum throughput of the users in the cell, while our approach leads to an overall spectral efficiency increase.

\emph{ Main Contributions:} One of the contributions of this paper is the introduction of a new paradigm for wireless video content dissemination with no additional infrastructure cost~\footnote{We first suggested the concept in a submission for the Intel VAWN funding program \cite{VAWN_2010} and in the conference version of this paper, \cite{golrezaei2012base}.}.
Our approach is based on caching popular video files in mobile devices, where groups of mobile devices create large "virtual" caches, in which file duplication is avoided as much as possible. The stored files are transmitted, upon request, to a user requesting a particular file, using highly spectrally efficient D2D transmission.


As a second contribution, we provide an approximate analysis of such a system. The analysis is based on a subdivision of a macrocell into virtual clusters, such that one D2D link can be active within each cluster. The cluster size is a key parameter of the system, and can be controlled and optimized by the transmit power of the mobile terminals. We investigate two possible caching schemes, the first has a central control of all devices (and a-priori knowledge of where devices will be when files are requested); the second is based on random caching where each device stores files without any central control. We also provide a simulation evaluation section based on real-world video popularity distributions taken from \cite{tracedata}. From these results, we find that improvements of the video throughput by one to two orders of magnitude are possible.

The remainder of the paper is organized as follows: Section \ref{sec_arch} describes our new architecture and its mathematical model. Section \ref{sec_model} analyzes the optimal placement of files in the cache. Section \ref{sec_optimal} derives the optimal cluster size (transmission distance between nodes). Section \ref{sec_exp} analyzes the overall performance of the scheme, and Section \ref{sec_conc} provides conclusions and an outlook for future studies.



\section{A new architecture for cellular video}\label{sec_arch}

Our new approach to solving the video bottleneck in cellular systems is based on the following two key observations: a large amount of video traffic is caused by a few, popular, files and storage is a quantity that increases faster than any other component in communications/processing systems:
\begin{itemize}
\item The popularity of video files is very unevenly distributed. "Viral" YouTube videos, movies that are newly available for rental, and reports from recent sports events are typical examples of highly popular videos, which account for a considerable percentage of all video traffic. In current cellular networks, the video is downloaded by each requesting user via the base station of the cellular network, which wastes precious spectral resources.  On the other hand, the redundancy cannot be exploited by a simple broadcasting scheme, because consumers want to view video {\em on demand}, and not be bound by predetermined starting times.
\item Storage space is the fastest-growing quantity on communication devices, outpacing the growth in data rate, and even Moore's law. In particular, recent years have seen an enormous proliferation of smartphones and tablets that have anywhere between $10$ and $64$ GByte of storage (not to mention the $500$ GByte on typical laptop harddisks). This storage is commonly under-utilized.
\end{itemize}
Users request video files according to an (empirical) distribution that reflects the popularities of the files. This distribution changes slowly (over the course of days or weeks), reflecting, e.g., the popularity of particular YouTube videos, recently released movies, etc. Thus, the popularity of files can be learned and predicted. However, caching of the most popular files at each wireless device is not effective, since the storage space at one particular node is insufficient to store more than a few popular files. We suggest that groups of mobile devices collaborate to exchange files via D2D communications. We can say that clusters of collaborating devices "pool" their caching resources to provide a "central virtual cache" (CVC).

\begin{figure}[htb]
    \centerline{\includegraphics[width=8.2cm]{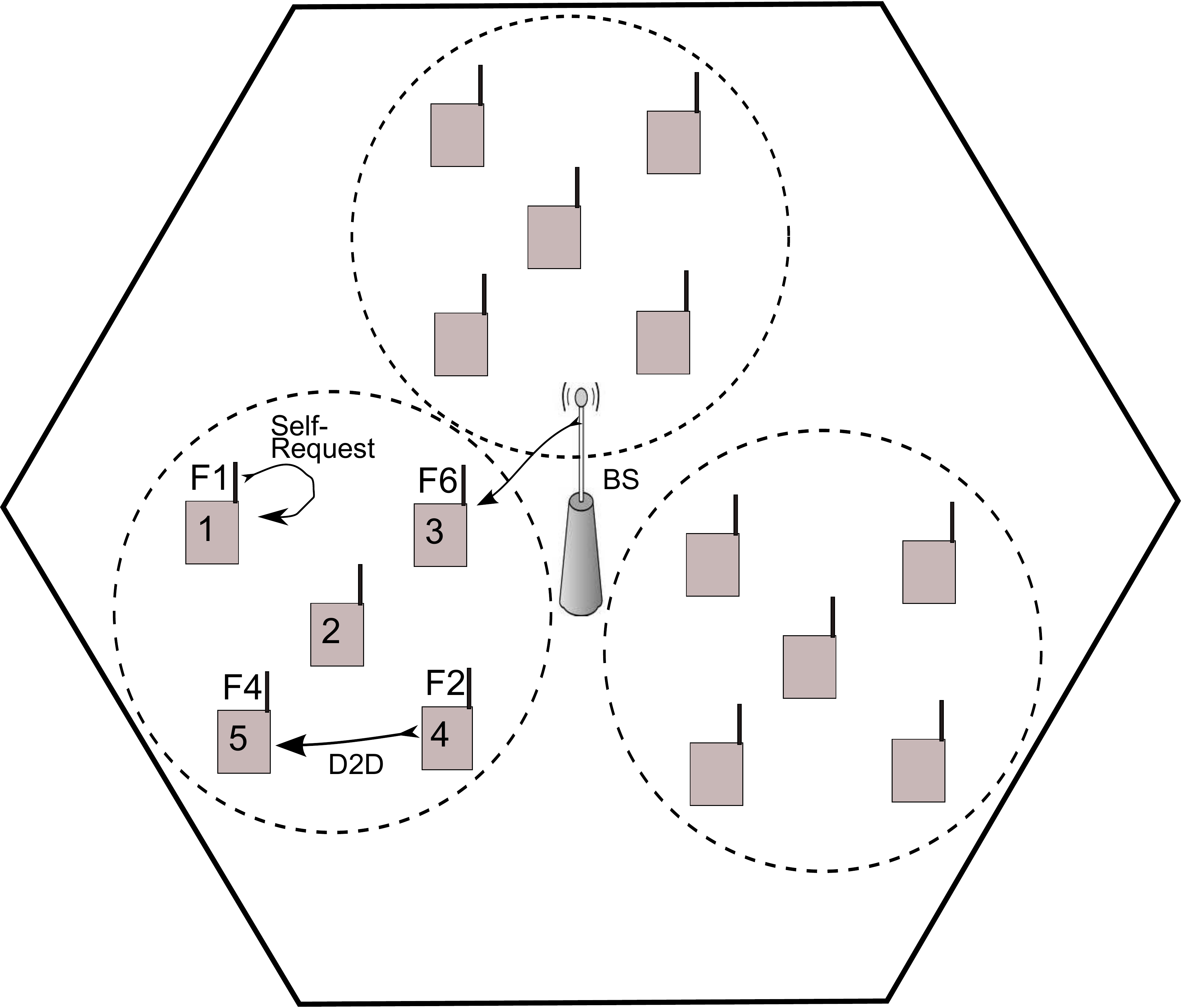}}
    \caption{An example of the single-cell layout.}
    \label{single_cell}
    \end{figure}

{ Figure \ref{single_cell} illustrates the overall architecture of a single cell with a BS and multiple users. Dashed circle shows a group of users that communicate with each. 
We divide the macrocell into groups or "clusters" of nodes (indicated by dashed circles)
that can exchange information with each other, constituting the virtual caches.
Each of the devices store files according to a specific caching rule - the choice of this rule will be discussed below.
For example, without loss of generality, we assumed that user $i$ stores file $F_i$.
Users request files from the virtual cache. If the video file is found in the part of the virtual cache that is located on the users own device,
(\textit{e.g.}, user 1 which requests $F_1$),
then the file is available immediately (we call this case a "self-request" - it poses the least demand on the wireless resources). If the file is found in the storage of another device in the group, i.e., another part of the virtual cache, then the file is sent through D2D communications to the requesting user,
for example, in Figure \ref{single_cell} user 5 requests file $F_4$ which is stored in cache of user 4, thus a D2D communication will be established from user 4 to user 5.
Such D2D communications occur over a short distance, and are highly spectrally efficient: (i) short distance implies high SNR, so that a high rate can be achieved for such a link, and (ii) short-distance communications allows high spatial reuse, which implies high area spectral efficiency.

We assume in this paper that all nodes use the same transmission power (i.e., there is no power control), but this common power can be optimized. The power determines the feasible communication range, and thus the cluster size. Finding its optimal value involves a trade-off between two counter-acting effects: (i) a small cluster size means a high spatial reuse, i.e., more D2D links can be operated simultaneously within one cell, (ii) a larger cluster size increases the probability that a user actually finds the file it requests in its vicinity.
Thus a D2D link can be meaningfully established and the corresponding cluster is called active.

The D2D communications is controlled by the BS: when a user requests a certain video file, the BS can inform it about the nearest device that has the file stored, and the file is then transmitted from that device via a D2D link. Since the BS has knowledge of all the links that can be scheduled, as well as the channel state information, it can optimize the frequency reuse between the devices. The distance between transmitting and receiving device is much shorter than between device and base station. Thus multiple D2D links can be operated on the same time/frequency resources within one cell.

If a requesting device does not find the file in its neighborhood or in its own cache,
it obtains the file in the traditional manner from the BS. We assume that the D2D communication does not interfere with communication between the BS and users. This assumption is justified if the D2D communications occur in a separate frequency band (\textit{e.g.}, WiFi). For the D2D throughput, we henceforth do not need to consider explicitly the BS and its associated communications.

\section{Model and Setup}\label{sec_model}

Assume a cellular network where each cell/base station (BS) serves $n$ users. For simplicity we assume that the cells are square, and we neglect inter-cell interference, so that we can consider one cell in isolation. Users are distributed uniformly in the cell, and can communicate with each other through D2D communications, as well as with the BS through traditional cellular (uplink/downlink) communication. Each user device has a cache in which video files can be stored. For rotational simplicity, we assume that each user stores exactly one file, though generalization to multiple files per user is trivial. The BS is aware of the stored files and channel state information of the users and controls the D2D communications.

For our theoretical derivations, we use simplified description of the physical layer, namely the protocol model, widely used in networking: two users can communicate if their physical distance is smaller than some collaboration distance $r$ ~\cite{gupta2000capacity,RGGbook}.
The maximum allowable distance for D2D communication $r$ is determined by the  power level for each transmission. Conversely, we assume that (given that node $u$ wants to transmit to node $v$) any transmission within range $r$ from receiver $v$ can introduce interference for the $u-v$ transmission. Thus, they cannot be activated simultaneously.

We assume that all D2D links share the same time-frequency transmission resource within one cell area. Multiple transmissions on those resources are possible since the distance between requesting users and users with the stored file will typically be small.  Furthermore, there should be no interference of a transmission by others on an active D2D link. To achieve this, the cell is divided into smaller areas called cluster, see Figure \ref{fig:layout}. All clusters  are assumed to be square with equal area. As can be seen from  figure \ref{fig:layout}, the cell side is normalized to $1$ and cluster side  is equal to $r$. We call $r$ \emph{the collaboration distance}.
Users in each cluster can communicate with each other locally; however to avoid intra-cluster interference, only one such communication per cluster is allowed. Interference to other clusters is assumed to be negligible, or can be made negligible through an appropriate frequency reuse scheme.
Clearly the physical-layer communication model is oversimplified, as it does not account for the inter-cluster interference, the fact that the pathloss coefficients might be different in different parts of the cell, and fading. However, our results indicate that it captures a fundamental tradeoff in D2D collaboration and as shown in our simulations, can give very high gains. Furthermore, we will demonstrate in Section \ref{sec_exp} the results that are closely aligned with those of a more sophisticated channel and interference model. In other words, when we will take into account pathless, interference and shadowing, the optimal collaboration distance that maximizes the rate is very close to that in the simplified physical layer model employed in this section.

We assume that users may request files from a ``library'' of size $m$.
Each user requests a file from the library independently, according to a given popularity distribution.
Based on numerous studies, Zipf distributions have been established as good models to the measured popularity of video files \cite{zipf,tracedata}. Under this model, the frequency of the $i$-th popular file, denoted by $f_i$, is inversely proportional to its rank:
 \begin{equation}\label{zipf}
 f_i=\frac{{\frac{1}{{{i^{\gamma_r} }}}}}{{\sum\limits_{j = 1}^m {\frac{1}{{{j^{\gamma_r} }}}} }},\,\,\ 1\leq i\leq m.
 \end{equation}
 The Zipf exponent $\gamma_r$ characterizes the distribution by controlling the relative popularity of files.
 Larger $\gamma_r$ exponents correspond to higher content reuse, \textit{i.e.,} the first few popular files account for the majority of requests.

 Thus, in order to enable D2D communication it is not sufficient that the distance between two users be less than $r$; users should also find their desired files locally in the CVC of their cluster, i.e., stored by one of the devices in the cluster. A link between
two users will be called \emph{potentially active} if one requests a file that the other is caching.
 Therefore, the probability of D2D collaboration opportunities depends on what is stored and requested by the users.  The decision of what to store can be taken in a centralized or distributed way, called deterministic and random caching, respectively.
 \begin{itemize}
\item {Deterministic caching} gives the optimal cache configuration. However, it is only feasible if the location of the nodes and the Channel State Information (CSI) is known a priori, and remains constant between the filling of the cache and the actual file transmission. The filling of the cache can take many hours or days, such an assignment is really only feasible for fixed wireless devices. Thus we consider the deterministic caching mostly as an upper bound for the achievable performance. Furthermore, finding the optimal deterministic file assignment is not trivial. In a general random geometric graph, it can be shown \cite{IT_D2D_paper} that finding the optimal deterministic file assignment is NP-Complete even when users are static and interference is ignored. However, in 
a clustering-based caching, we can find a simple method for centralized file placement. Given that there are $k$ users in a virtual cluster, each of $k$ users should store one of the $k$ most popular files without repetition.
\item{Random caching} can be implemented when users are highly mobile. Each user caches files at random and independently, according to a caching distribution. Specifically, we make a heuristic choice for this caching distribution, namely a Zipf distribution with exponent $\gamma_c$ \footnote{For a slightly different deployment model, and the limit of large number of nodes, \cite{ji2013optimal} recently derived a "waterfilling-type" optimum caching distribution.}. The exponent of caching distribution is one of our decision variables which is not necessarily equal to $\gamma_r$. Thus, in this strategy we find the optimal $r$ and $\gamma_c$ to maximize the number of D2D links. Since the BS does not influence which user stores which file, and the presence of users in a particular cluster is random, there may be overlap of the cache content at different users. Such a duplication makes the CVC less efficient. Still, we will see in Section \ref{sec_exp} that the performance loss compared to the deterministic caching is small.
\end{itemize}

   \begin{figure}
\centering
\centering \includegraphics[width=6cm]{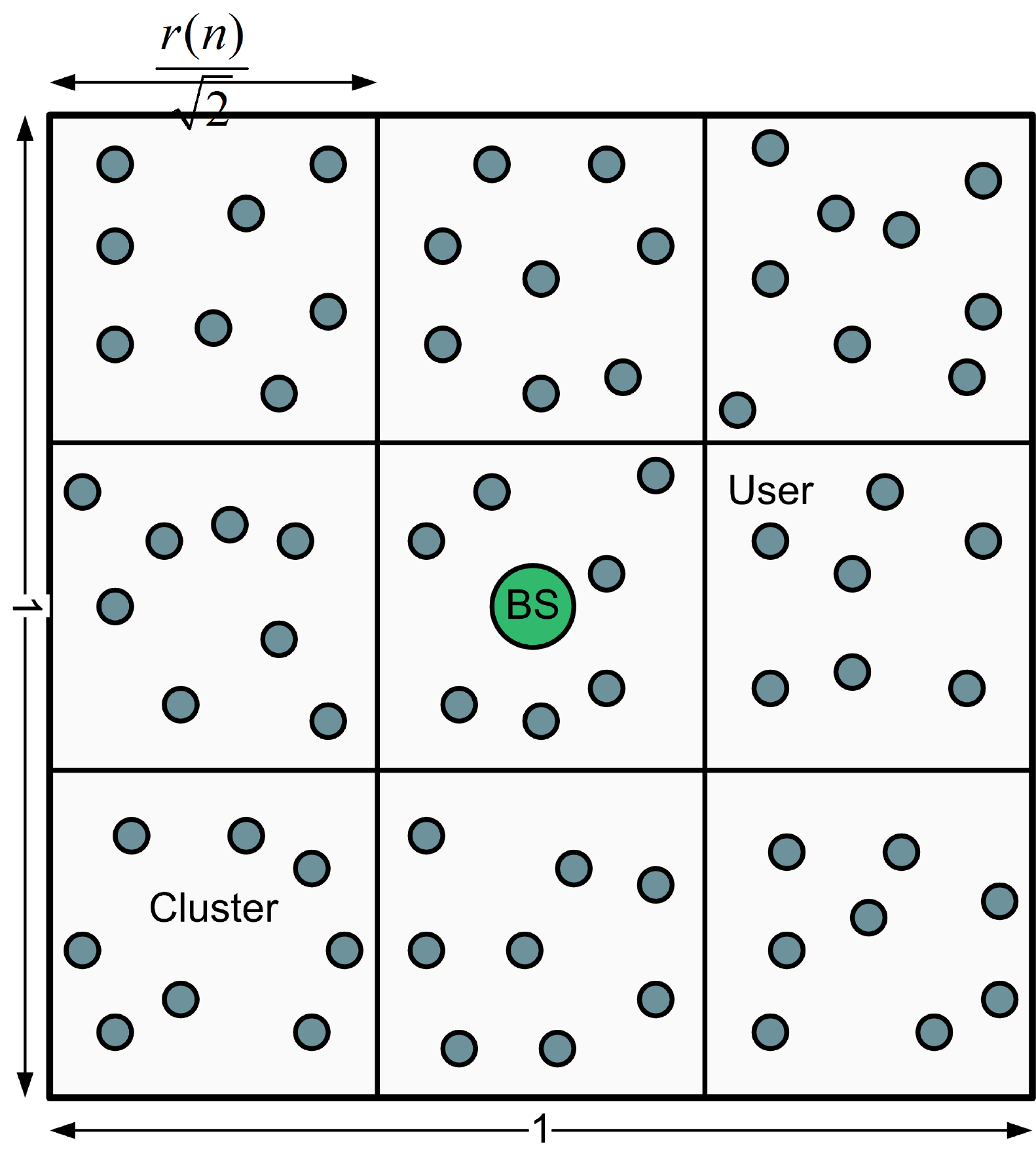}
\caption{a)~Dividing cell into virtual clusters. b)~In the worst case, a good cluster can block at most $16$ clusters. In the dashed circle, receiving is not possible and in the solid circle, transmission is not allowed.}
\label{fig:layout}
\end{figure}

\section{Finding the optimal collaboration distance}\label{sec_optimal}

\subsection{Deterministic caching}
In this section, we find the optimal $r$ for given values of size of the library $m$ and the number of users in the cell $n$ when the users store files based on the deterministic caching strategy. In this strategy if there are $k$ users in a virtual clusters, each of them should store one of the $k$ most popular files without repetition.

In the subsequent formulation we assume that two neighbor clusters can have D2D links simultaneously and we ignore the inter cluster interference, so the number of D2D links is equal to the number of active clusters.\footnote{If we use guard distances to avoid inter-cluster interference, the number of active clusters is scaled by this constant "frequency reuse factor" \cite{Molisch_book_2011}, Chapter 17. } In our simulation results, we justify this assumption by  showing that the effect of interference is negligible.

We define a binary random variable $a_j$ for cluster $j$ such that $a_{j}$ is equal to $1$ if the cluster $j$ is  active; otherwise, it is equal to $0$. The total number of active clusters in the cell, denoted by $A$, equals to:\\
\begin{equation}\label{A}
A=\sum\limits_j {a_j}.
\end{equation}
Since $a_j$ is a binary random variable, the expectation of $a_j$ is the probability that cluster $j$ is active, i.e., D2D communication takes place in cluster $j$. Since users are uniformly distributed in the cell and all clusters have equal area, the expectation of $a_{j}$, denoted by $E[a_j]$, does not depend on $j$, i.e., $E[a_j]=E[a]$ for any $j$ where $E[a]$ is the probability that any cluster is active. Thus, from (\ref{A}), the expected number of active clusters is given by:\\
\begin{equation}\label{E[y]}
E[A] = \sum\limits_j {E[{a_j}]}  = \frac{2}{{{r^2}}}E[a],
\end{equation}
where $\frac{2}{r^2}$ is the number of clusters in the cell. The probability that a cluster is active depends on number of users in the cluster which is denoted by $K$. Therefore, $E[a]$ can be written as:\\
\begin{equation}\label{E[X]}
E[a] = \sum\limits_{k = 0}^n {E[a|K = k]\Pr [K = k]},
\end{equation}
where  $E[a|K = k]$ is the probability that a cluster is active provided that there are $k$ users in the cluster. $\Pr [K = k]$ is the probability that the number of users in the cluster is $k$.
The number of users in the cluster is a binomial random variable with parameters $n$ and $\frac{r^2}{2}$, i.e., $K=B(n,\frac{r^2}{2})$.
Note $\frac{r^2}{2}$ is the ratio of the cluster area to the cell area.
The probability that there are $k$ users in the cluster equals to:\\
\begin{equation}\label{pr[K=k]}
\Pr [K = k] = \left( \begin{array}{l}
n\\
k
\end{array} \right){(r^2/2)^k}{(1 - r^2/2)^{n - k}},
\end{equation}
 where $
 \left( \begin{array}{l}
 n\\
 k
 \end{array} \right) = \frac{{n!}}{{(n - k)!k!}}$.
The expectation $E[a|K=k]$, defined in (\ref{E[X]}), is the complement of the probability that no D2D communication takes place in the cluster. D2D communication is possible if at least one of $k$ users in the cluster can access to its requested file in the cache of other users. 
Thus, $E[a|K=k]$ can be written as:\\
 \begin{equation}\label{E[a|K=k] and independence}
         E[a|K = k]=1-\Pr[u_1=1 \cap u_2=1 \cap ... \cap u_k=1].
         \end{equation}
The $u_i$ for $i=1,...,k$ is a binary random variable that is $1$ if the user $i$ cannot find its requested file in the neighbors' caches in the cluster, i.e., in the CVC excluding the file in $i$th user's cache.

According to our assumptions in this section, given $k$, the CVC in the cluster is  {\em deterministic}. Since users' requests are independent from each other, the random variables  $u_i$ and $u_j$ for $i\neq j$ are also independent.
     Thus, we can simplify Eq. (\ref{E[a|K=k] and independence}) as:\\
        \begin{equation}\label{first_E[a|k]}
             E[a|K = k]=1 - \prod\limits_{i = 1}^k {\Pr [{u_i} = 1]}.
        \end{equation}
        Without loss of generality, we assume that user $i$ caches the $i$th most popular file. Note furthermore the possibility of {\em self-requests}, i.e., a user might find the file it requests in its own cache. In this case clearly no D2D communication will be activated by this user. Accounting for these self-requests,
        \begin{equation}\label{ui}
        {\Pr [{u_i} = 1]}=1-(P_{CVC}(k)-f_{i}),
        \end{equation}
        where $f_i$ is the request probability of the $i$th popular file and is given in (\ref{zipf}). $P_{CVC}(k)$ is the probability of hitting the CVC and is given by
             \begin{equation}\label{main_p_cache}
             P_{CVC}(k)=\sum\limits_{i = 1}^k {{f_i}},\,\,\, 1\le k\leq m,
             \end{equation}
              It is obvious that $P_{CVC}(k)$ for $k>m$ is equal to $1$. From (\ref{first_E[a|k]}) and (\ref{ui}), we find:\\
             \begin{equation}\label{E[X|K=k]}
                       E[a|K = k]=1 - \prod_{i=1}^{k}(1-(P_{CVC}(k)-f_i)).
             \end{equation}
            Substituting $E[a|K=k]$ in $(\ref{E[X]})$, we get:\\
  \begin{equation}\label{E[x]_2}
  E[a] = \sum\limits_{k = 0}^n {\left( 1 - \prod_{i=1}^{k}(1-(P_{CVC}(k)-f_i))\right)\Pr [K = k]},
  \end{equation}
  where $Pr[K=k]$ is given in (\ref{pr[K=k]}).
 From (\ref{E[y]}), (\ref{E[X]}), and (\ref{E[x]_2}), the expected number of active clusters can be written as:\\
\begin{subequations}\label{E[y]2}
\begin{align}
E[A] &= \frac{2}{{{r^2}}}\sum\limits_{k = 0}^n {E[a|K = k]\Pr [K = k]} \label{E[y]2a}\\
  &= \frac{2}{{{r^2}}}\sum\limits_{k = 0}^n {\left( 1 - \prod_{i=1}^{k}(1-(P_{CVC}(k)-f_i))\right)\Pr [K = k]}.
\end{align}
\end{subequations}
  Notice that $\Pr[K=k]$ is a function of $r$. Thus, $E[A]$ is a function of an unknown variable $r$. To find the maximum expected number of active clusters or equivalently the maximum average spectral efficiency, we should take a derivative of $E[A]$ with respect to $r$. While finding $r_{opt}$ analytically in closed form does not seem feasible,  numerical solutions are possible with very low effort, as we require a root search of a function of one variable within the interval $0<r<1$.

In this optimization problem, the self-requests do not play any role. However, some users can get files with zero delay because of self-requests. This will suggest to consider  an alternative criterion, minimization of the download time as follows

  \begin{equation}
\mbox{minimize}\,\,\,\, (n-n_{self}-E[A])\omega_{BS}+E[A]\omega_{D2D},\label{min_delay}
\end{equation}
where $n_{self}=\frac{1}{r^2}\sum_{k=0}^{n}P_{CVC(k)}\Pr[K=k]$ is the average number of users that get their desired files from their own cache with zero delay. $\omega_{D2D}$ and $\omega_{BS}$ are the average download time through D2D communications and the BS, respectively. Thus, the average total delay of all users equals:
 \[\bar{D}=(n-n_{self}-E[A])\omega_{BS}+E[A]\omega_{D2D}.\]\label{min_delay2}

\subsection{Random caching}
    In this strategy, we again assume that the BS divides the cell into virtual clusters with size $r$.
Our goal here is to maximize the average number of clusters, but now under the assumption of random caching. The first steps are similar to previous subsection.
The expected number of active clusters and the probability that a cluster is active  are given by (\ref{E[y]})  and (\ref{E[X]}). The probability that a cluster is active when there are $k$ users in the cluster is stated in (\ref{first_E[a|k]}). 

 However, for the next step, we cannot directly apply the approach used in the previous subsection, since with a random caching strategy, $u_i$ and $u_j$ for $i\neq j$ will be {\em dependent} random variables. This can be understood as follows: if user $i$ requests a file according to the some popularity  distribution and cannot find it in the CVC, it implies an increased probability that the current content of the
the CVC does not contain (some) popular files. So, if user $j$ requests a file independently from user $i$ but according to  the same popularity  distribution, it is more probable that user $j$ also cannot find its requested file in the CVC in the cluster.

 Assume now that users $\{1,2,\ldots,k\}$ in a cluster respectively store the files $H=[h_1, h_2, \ldots, h_k]$, then
          \begin{equation}\label{u2}
        {\Pr [{u_i} = 1]}=\sum_{H \in \{1,\ldots ,m\}^k}\big[1-(\sum_{h_j \in \tilde H}f_{h_j}-f_{h_i})\big] \Pr[H] ,
        \end{equation}
where $\tilde{ H}=\cup_{j=1}^k \{h_j\}$. Since users store files independently, the probability that users in the cluster store files $H$, $Pr[H]$,
     can be written as
     \begin{align}
     \Pr[H]=\prod_{j=1}^k p_{h_j}
     \end{align}
where $p_{h_j}$ is the probability that user $j$ stores file $h_j$, more precisely
 \begin{equation}\label{zipf2}
 p_{h_j}=\frac{{\frac{1}{{{{h_j}^{\gamma_c} }}}}}{{\sum\limits_{i = 1}^m {\frac{1}{{{i^{\gamma_c} }}}} }},\,\,\ 1\leq h_j\leq m.
 \end{equation}
 As stated before, we assume that users store files according to Zipf distribution with exponant $\gamma_c$.
Using (\ref{E[y]}), (\ref{E[X]}), (\ref{first_E[a|k]}) and (\ref{u2}), the expected number of active clusters is given by 
\begin{align}
\nonumber &E[A]
  = \frac{2}{r^2}\sum\limits_{k = 0}^n \Big[\Pr [K = k]\\
 &\times  \Big( 1 - \prod_{i=1}^{k}\sum_{H \in \{1,\ldots ,m\}^k}\big[1-(\sum_{h_j \in \tilde H}f_{h_j}-f_{h_i})\big] \Pr[H] \Big)\Big].
\end{align}
$E[A]$ is a function of $r$ and $\gamma_c$. The optimal $r$ and $\gamma_c$ can be found numerically using the above equation.
Note that the computation of this probability involves a summation over exponentially many terms. We estimate this using Monte Carlo simulations which asymptotically converge (by the Law of Large Numbers) to the correct probability. Obtaining error bounds on the rate of this convergence for a finite number of Monte Carlo iterations and how their number needs to scale in the size of the problem is an interesting theoretical question that we do not investigate in this paper. We empirically found that our estimates converged after $1000$ Monte Carlo iterations for the tested cases.


Similar to previous section, one can minimize the download time as follows
  \begin{equation} \label{max problem2}
\mbox{minimize}\,\,\,\,(n-n_{self}-E[A])\omega_{BS}+E[A]\omega_{D2D} 
\end{equation}
where $\omega_{D2D}$ and $\omega_{BS}$ are defined in (\ref{min_delay}).
Given that users stores files $F$, $n_{self}=\sum_{f_{h_j} \in F}f_{h_j}$ where $f_{h_j}$ is the probability that agent $j$ requested its own stored file $h_j$. Thus, we have \[n_{self}=\sum_{k}\Pr[K=k]\sum_{H \in \{1,\ldots ,m\}^k} \Pr[H] \sum_{h_j \in H}f_{h_j} \]


\section{Evalutaions by Computer Experiments}\label{sec_exp}

In this section, we provide some numerical results to investigate the effects of system parameters on the optimal collaboration distance for deterministic and random caching strategies. In all figures except those considering the effect of the parameters, the number of users in the cell is $500$ and the number of files is $1000$.
 \subsection{Deterministic caching}
 In this subsection the performance of the deterministic caching is studied. First, we want to find the optimal collaboration distance which maximizes the average number of active clusters. In this optimization problem, we face with a trade-off between two factors: First, the smaller the collaboration distance, the higher frequency reuse. Second, the larger the collaboration distance, the higher probability of finding the requested files stored in the same cluster.

Figure \ref{fig:EA_r_dgamma} shows the average number of active clusters versus the collaboration distance $r$, where the interference between clusters are ignored.
We can observe that $r_{opt}$ decreases as $\gamma_r$ increases. This agrees with the following intuition: the larger $\gamma_r$, the higher the probability that a (popular) file can be found very close to the requesting device, and the smaller the clusters can be made.

 \begin{figure}[htb]
    \centerline{\includegraphics[width=14cm]{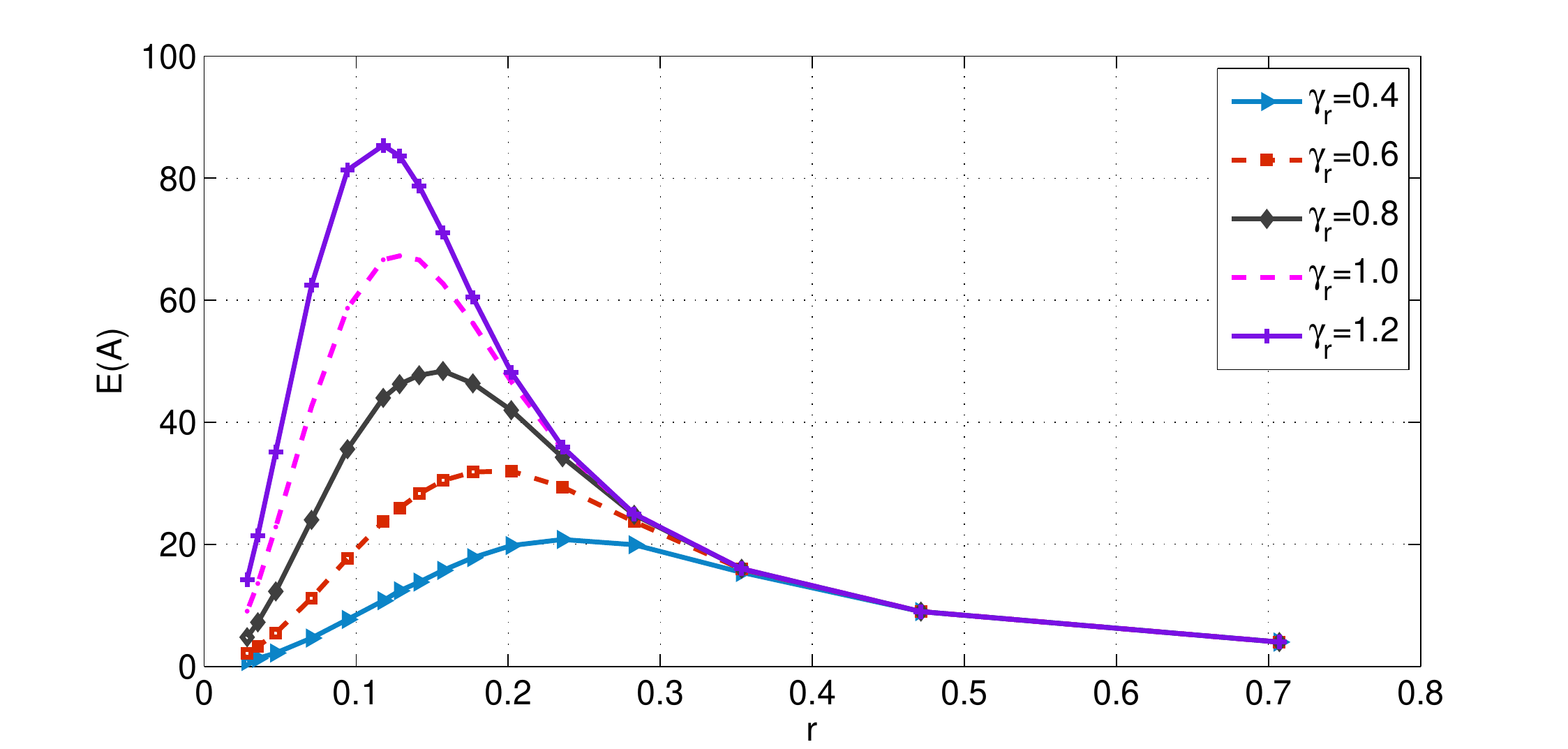}}
    \caption{The average number of active clusters versus the collaboration distance for deterministic caching with $n=500$ and $m=1000$.}
    \label{fig:EA_r_dgamma}
    \end{figure}

Next we study the effect of our simplifying assumptions. We analyze by simulations a system that does take into account the inter-cluster interference, as well as pathless and shadowing. In this system, we aim to maximize the total average rate (note that maximizing average rate and maximizing number of active clusters might be different in such a more elaborate setting, while it is the same for the simplified system studied up to now). To provide a realistic channel model, we assume a standard power-law pathloss with path-loss exponent $2.6$.  Super-imposed on that pathloss is log-normal shadowing with $\sigma=4dB$, which is assumed to be independent for all links. We do not include small-scale fading, assuming that it has been averaged out through (frequency-, time-, or antenna-) diversity. In order to include interference, we compute the optimal links, SIRs, and associated rates, as follows: in each cluster we assign files according to the deterministic distribution to the nodes, and generate file requests at random according to the Zipf distribution. We then establish all potential links in all the clusters. We now select active links from the set of potential links according to an iterative procedure: in a first round, each cluster schedules the links that has the highest received power. We next compute the interference computed by the scheduled transmitters at all receivers of potential links in the neighboring cells, as well as the SIR. If the scheduled link in one cell has a worse SIR than another link in that cell, we change the active link (which of course changes the interference in the neighboring cells). This procedure is repeated until no links change anymore.

The total average rate versus $r$ is shown in figure \ref{fig:rate_r_dgamma}.  Similar to the previous figure, the optimal collaboration maximizing the average rate, defined by $r_{opt}^R$, decreases as $\gamma_r$ increases. Comparing figures \ref{fig:EA_r_dgamma} and \ref{fig:rate_r_dgamma}, we can conclude that the optimal collaboration distance is slightly smaller when we take the interference into account.
Intuitively, a smaller cluster size means that we have higher frequency reuse, but the probability of having empty clusters, which serve as a protection region for the interference, is higher. Therefore, in the presence of interference, a small cluster size is better, \textit{i.e.}, $r_{opt}^R \leq r_{opt}$. Yet, we also observe that our (over)simplified model that forms the basis of our analytical treatment provides an optimal cluster size that is remarkably close to the optimal cluster size as obtained from the simulated rate optimization.

    \begin{figure}[htb]
    \centerline{\includegraphics[width=14cm]{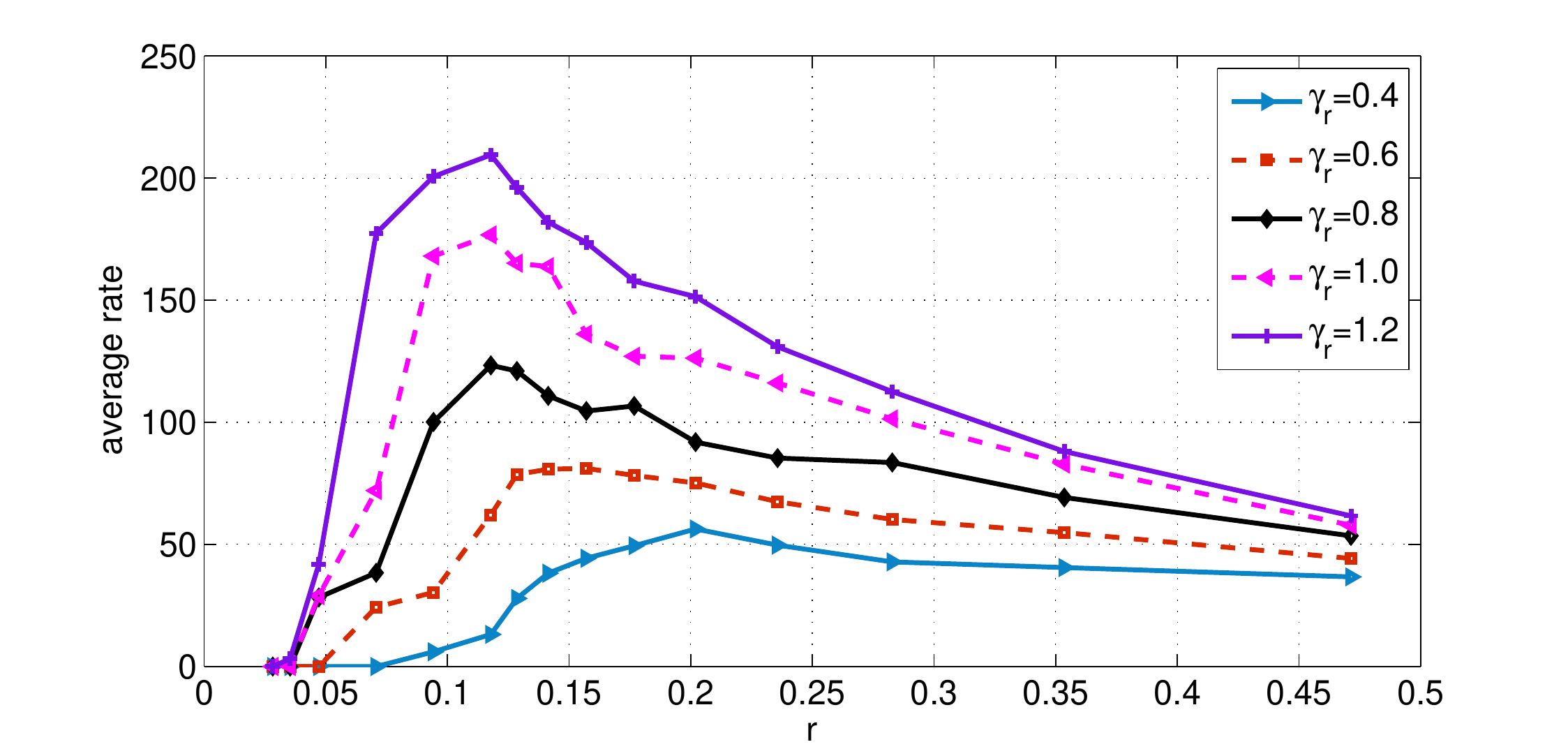}}
    \caption{The average rate versus the collaboration distance for deterministic caching with $n=500$ and $m=1000$.}
    \label{fig:rate_r_dgamma}
    \end{figure}


 From now on, we just present the simulation results for the average number of active clusters, but similar explanations are valid for the average rate. In the remaining figures except those we want to consider the effect of $\gamma_r$, we set it to $0.6$; this value is based on a study conducted on the University of MassachusettsÕ Amherst campus in $2008$  \cite{tracedata}.

 The effects of changing the popularity Zipf distribution exponent are investigated in Fig. \ref{fig:maxEA_gamma}. This figure shows the total number of clusters at $r_{opt}$, called optimal total number of clusters (which is simply $2/r_{opt}^2$), and $E(A)$ at $r_{opt}$, called the optimal average number of {\em active} clusters, versus $\gamma_r$. We can observe from the figure that by increasing  $\gamma_r$, the optimal total number of clusters and the optimal $E[A]$ increases. For the small $\gamma_r$, there is a little redundancy in the users' requests, i.e., the probability of finding files in the CVC is generally small. Thus, to increase the chance of having D2D communication within a cluster, the collaboration distance $r$ should increase. Hence, the  optimal average number of active clusters decreases for small $\gamma_r$, as already explained above. In short, more redundancy in video requests results in higher spectral efficiency.

    \begin{figure}[htb]
    \centerline{\includegraphics[width=14cm]{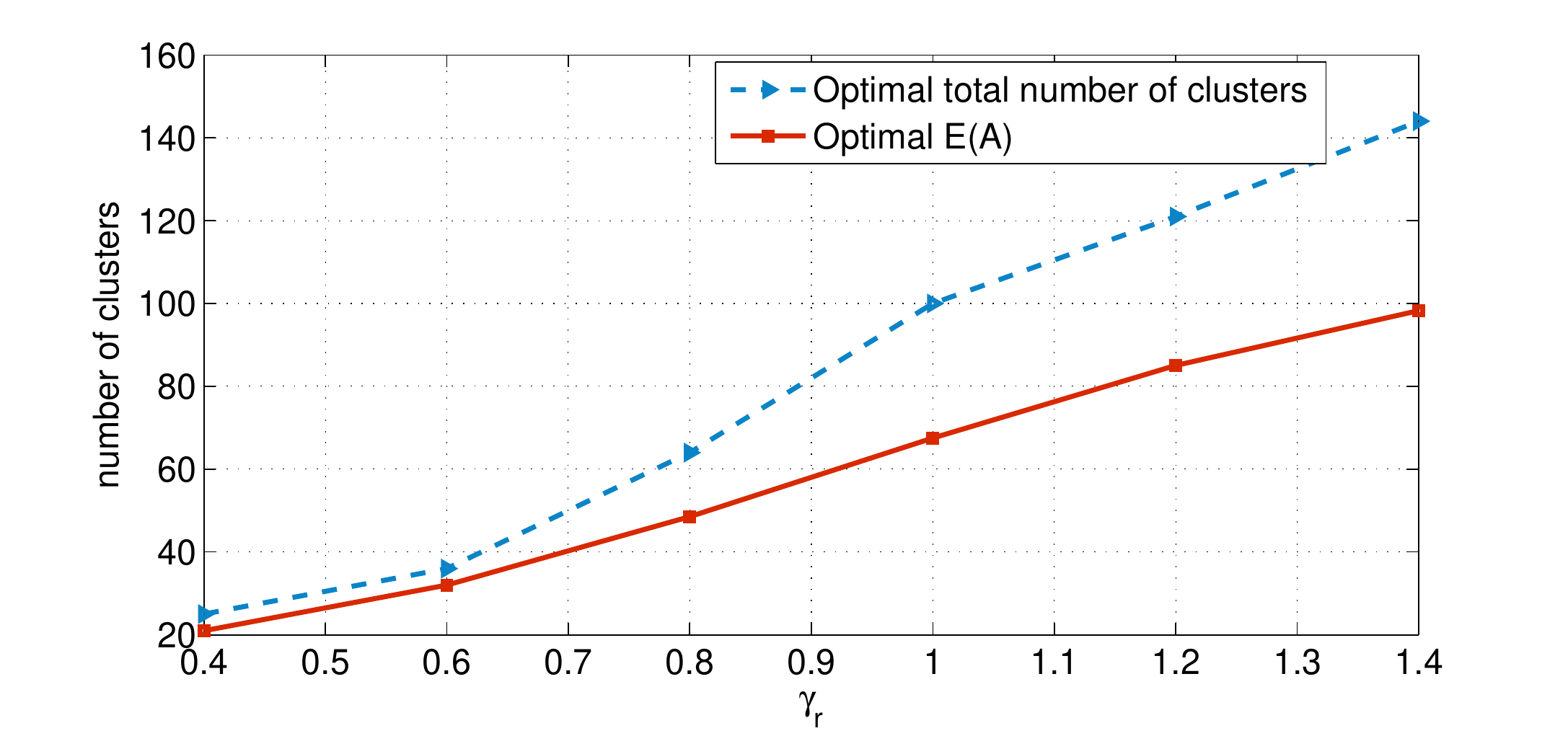}}
    \caption{The optimal total number of clusters and the optimal average number of active clusters versus $\gamma_r$ for deterministic caching with $n=500$ and $m=1000$.}
    \label{fig:maxEA_gamma}
    \end{figure}

In Fig. \ref{fig:ropt_gamma_fixed} the optimal collaboration distance achieved by two different optimizations is plotted versus $\gamma_r$. These two optimizations are based on maximizing the average number of active links, and minimizing the download time.
For larger $\gamma_r$, the $r_{opt}$ minimizing the download time is lower than that maximizing $E(A)$.
Because there is a higher probability that a popular file is stored locally. The smaller the cluster, the more often this popular file is cached. This will thus increase the probability of a self-request and decrease the cluster size (compared to the case of maximizing the average number of active links).

\begin{figure}[htb]
    \centerline{\includegraphics[width=14cm]{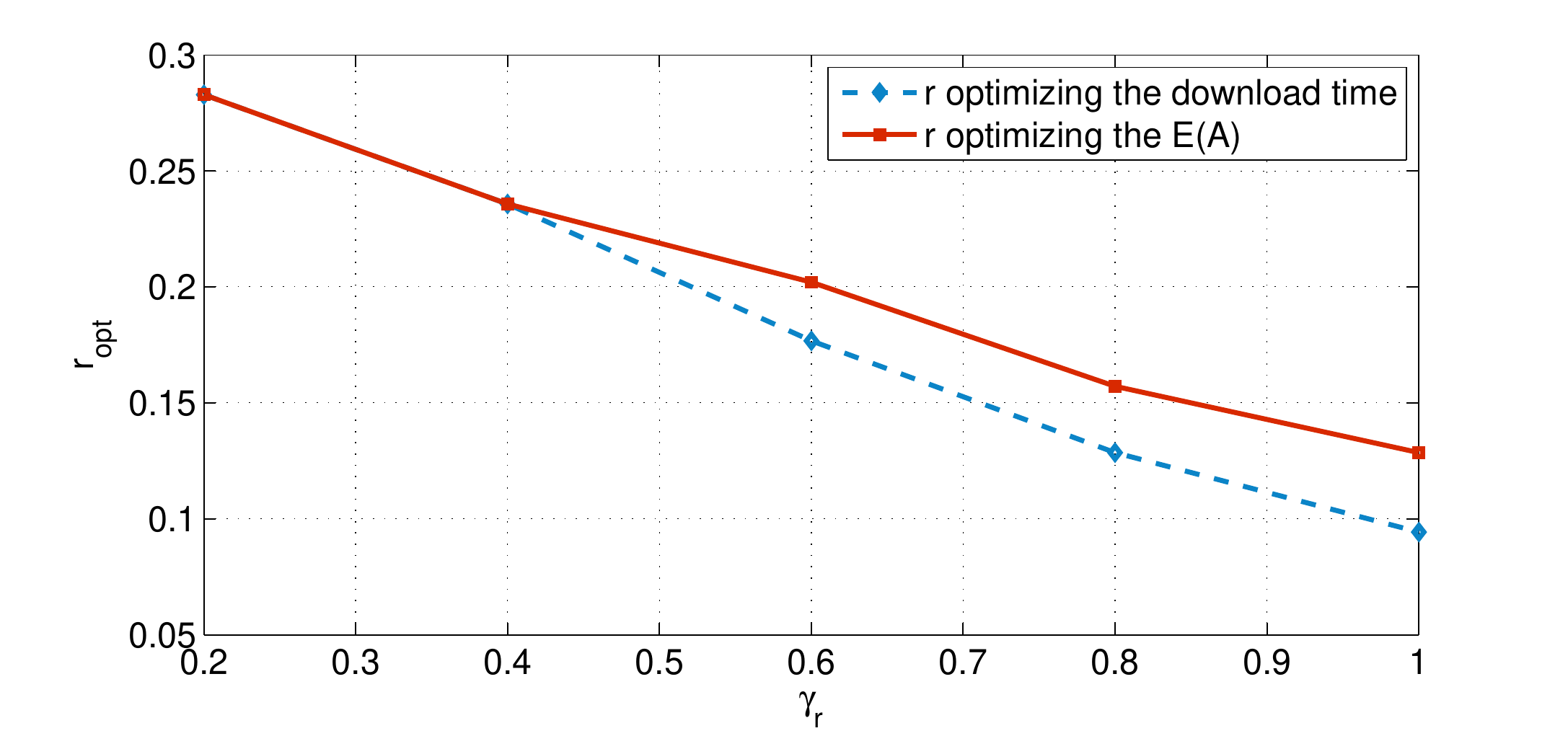}}
    \caption{The optimal collaboration distance versus versus $\gamma_r$ for deterministic caching with $n=500$ and $m=1000$.}
    \label{fig:ropt_gamma_fixed}
    \end{figure}

      Figure \ref{fig:maxEA_m} shows the effects of the size of the library on the optimal total number of clusters and the optimal average number of D2D links.

  As the size of the library increases, users request from a larger set of files and there is more diversity in requests. The probability that these diverse requests matches with the files stored in the CVC in the cluster decreases. So, by increasing $m$, users want to be surrounded by more neighbors in the clusters or equivalently to access to more files locally from the CVC. Hence, as it can be seen from figures, by increasing $m$,  the optimal $r$ increases and the optimal average number of active clusters decreases.

    \begin{figure}[htb]
    \centerline{\includegraphics[width=14cm]{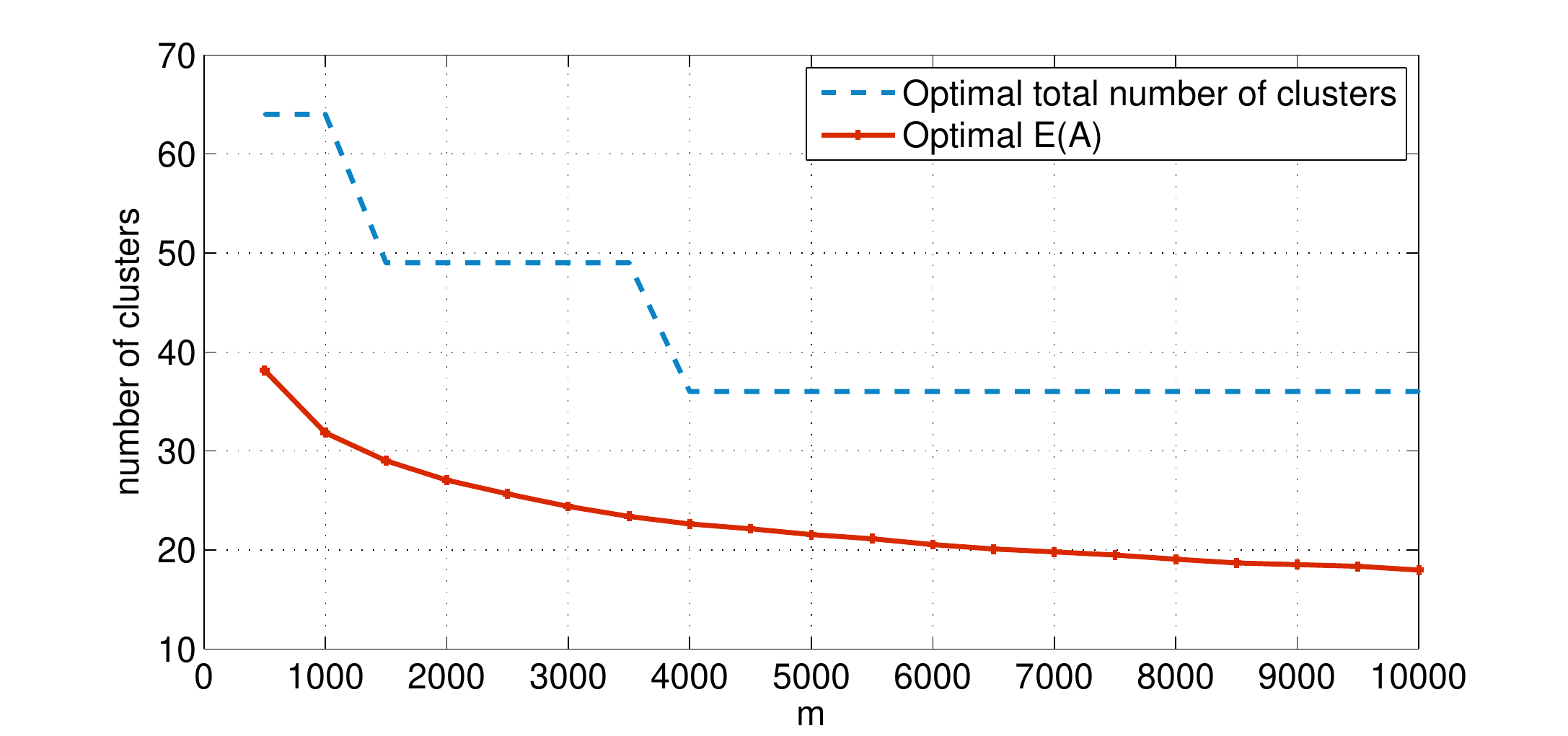}}
    \caption{The total number of clusters and the average number of active clusters versus the size of the library $m$ for deterministic caching with $\gamma_r=0.6$ and $n=500$.}
    \label{fig:maxEA_m}
    \end{figure}

 Fig. \ref{fig:maxEA_n_logx} and Fig. \ref{fig:ropt_rapprox} show the asymptotic behavior of the optimal collaboration distance and the number of active links as the number of users in the cell grows for two cases of $\gamma_r=0.6<1$ and $\gamma_r=1.4>1$. We assume there is a library of size $m=30 \sqrt{n}$ (for a justification of this type of proportionality, see \cite{golrezaei2012femtocaching}). For larger $n$, a user is surrounded by more neighbors given a fixed collaboration distance. In other words, users can find their desired files within a short distance which results in smaller optimal collaboration distance and larger number of active links.
  In our previous work, we have shown that for $\gamma_r<1$ (resp., $\gamma_r>1$), the optimal collaboration distance is $r_{opt}(n)=\Theta(\sqrt{\frac{m^{\eta}}{n}})$\footnote{We use the standard Landau notation: $f(n)=O(g(n))$ and $f(n)=\Omega(g(n))$ respectively denote $|f(n)|\leq c_1 g(n)$ and $|f(n)|\geq c_2 g(n)$ for some constants $c_1,c_2$. $f(n)=\Theta(g(n))$, stands for $f(n)=O(g(n))$ and $f(n)=\Omega(g(n))$.} (reps. $r_{opt}(n)=\Theta(\sqrt{\frac{1}{n}})$ ) and the number of active links scales with $\frac{n}{m^{\eta}}$ (resp., $n$) where $\eta=\frac{1-\gamma_r}{2-\gamma_r}$.
  Figure  \ref{fig:maxEA_n_logx} and \ref{fig:ropt_rapprox} confirm the asymptotic behavior of $r_{opt}(n)$ and $E[A]$.

\begin{figure}[htb]
    \centerline{\includegraphics[width=14cm]{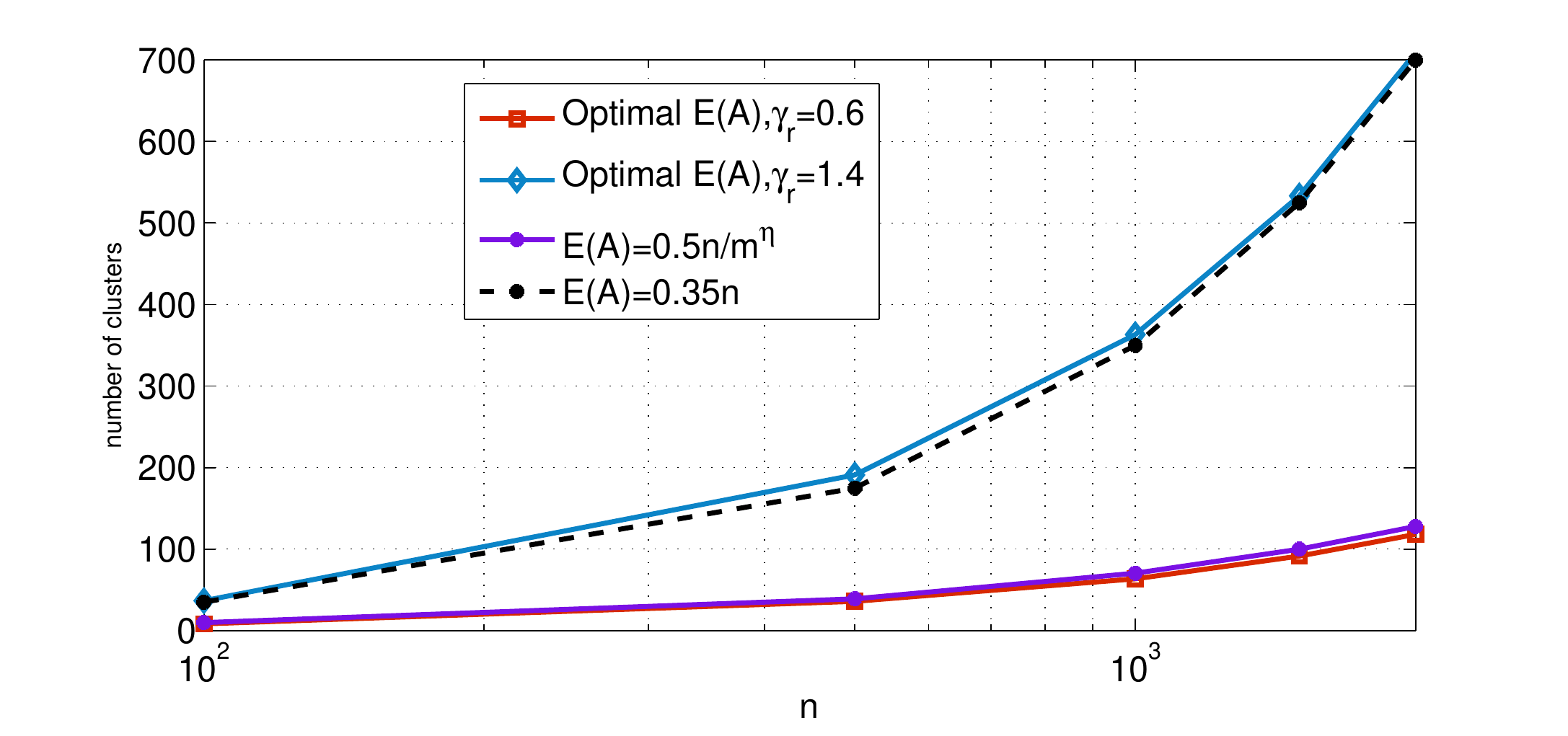}}
    \caption{The optimal average number of active clusters versus the number of users $n$ for deterministic caching with $\gamma_r=0.6,1.4$ and $m=30 \sqrt{n}$. $E_1=0.5\frac{n}{m^{\eta}} $ and $E_2=0.35n$. }
    \label{fig:maxEA_n_logx}
    \end{figure}

\begin{figure}[htb]
    \centerline{\includegraphics[width=14cm]{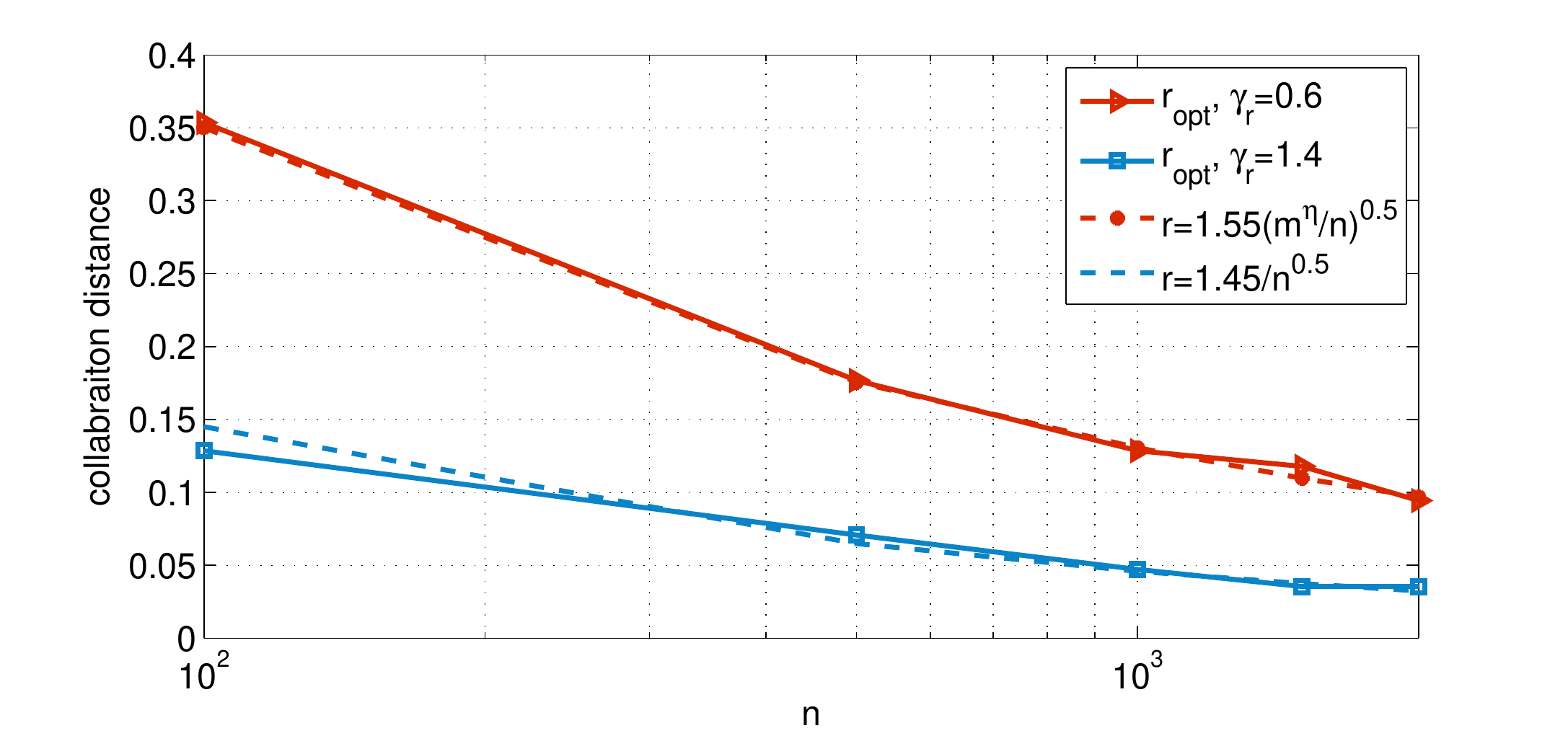}}
    \caption{The optimal collaboration distance versus the number of users $n$ for deterministic caching with $\gamma_r=0.6,1.4$ and $m=30 \sqrt{n}$. $r_1=1.55\sqrt{\frac{m^{\eta}}{n}} $ and $r_2=1.45\sqrt{\frac{1}{n}}$.}
    \label{fig:ropt_rapprox}
    \end{figure}

\subsection{Random caching}

In this subsection we study the performance of the random caching. Here we have to optimize two parameters: (i) is the collaboration distance $r$, and (ii) the exponent of caching distribution, $\gamma_c$.

Figures \ref{fig:3D_EA_g06} shows the average number of active clusters versus the collaboration distance $r$ and the exponent of caching distribution, $\gamma_c$. As this figure shows and Fig. \ref{fig:EA_r_dc} confirms, the optimal collaboration distance is equal to 0.2 and the optimal $\gamma_c$ is equal to 1.5.

	\begin{figure}[htb]
    \centerline{\includegraphics[width=14cm]{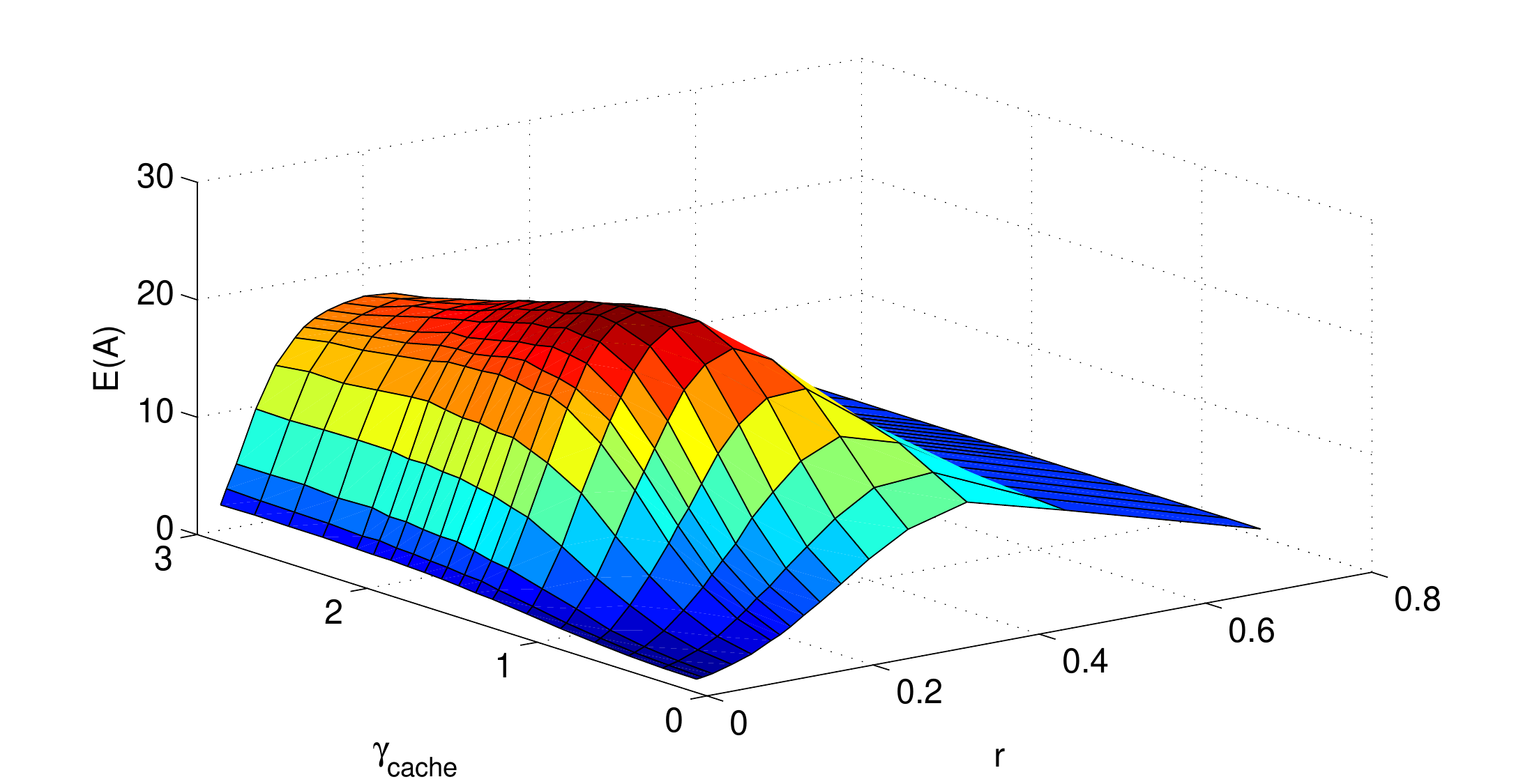}}
    \caption{The average number of active clusters versus $r$ and $\gamma_c$ for random caching with $\gamma_r=0.6$, $n=500$ and $m=1000$.}
    \label{fig:3D_EA_g06}
    \end{figure}

   \begin{figure}[htb]
    \centerline{\includegraphics[width=14cm]{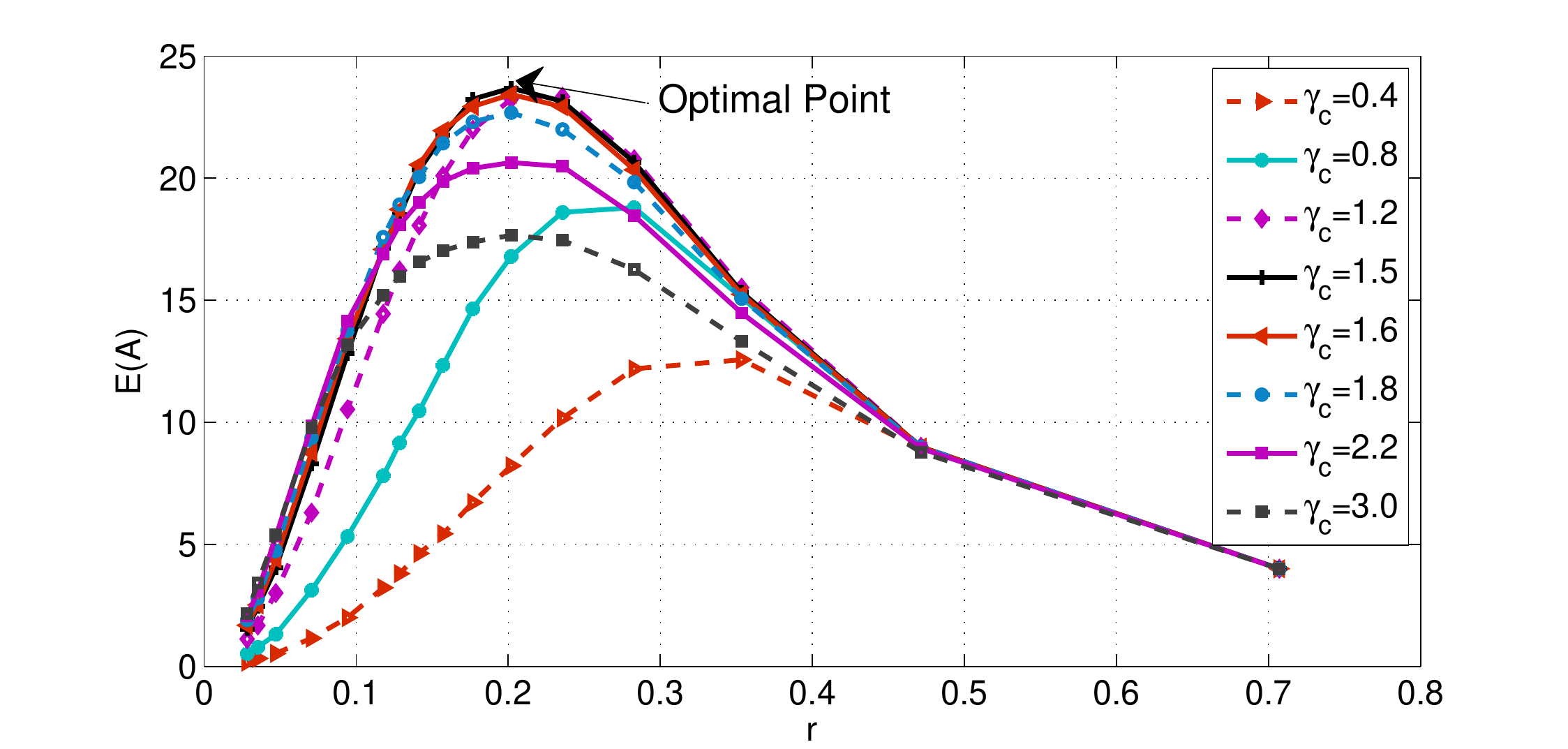}}
    \caption{The average number of active clusters versus $r$ for random caching for different $\gamma_c$, with $\gamma_r=0.6$, $n=500$ and $m=1000$.}
    \label{fig:EA_r_dc}
    \end{figure}

      The effect of system parameters on the optimal $r$ is similar to the deterministic caching strategy and due to space limitation we omit the corresponding figures.

  Figure \ref{fig:copt_gamma} shows how changing the request distribution exponent $\gamma_r$ affects $\gamma_{c,opt}$, the optimal exponent of caching distribution maximizing the average number of active clusters. 
 As it is obvious from the figure, by increasing $\gamma_r$, $\gamma_{c,opt}$ increases. For the larger $\gamma_r$, the first few popular files account for the majority of requests. 
 Therefore, to satisfy the users' requests, there is no need to cache less popular files and larger $\gamma_{c}$ could be optimal. Another observation is that the optimal $\gamma_c$ is not necessarily equal to $\gamma_r$ and it is much larger than it.

 \begin{figure}[htb]
    \centerline{\includegraphics[width=14cm]{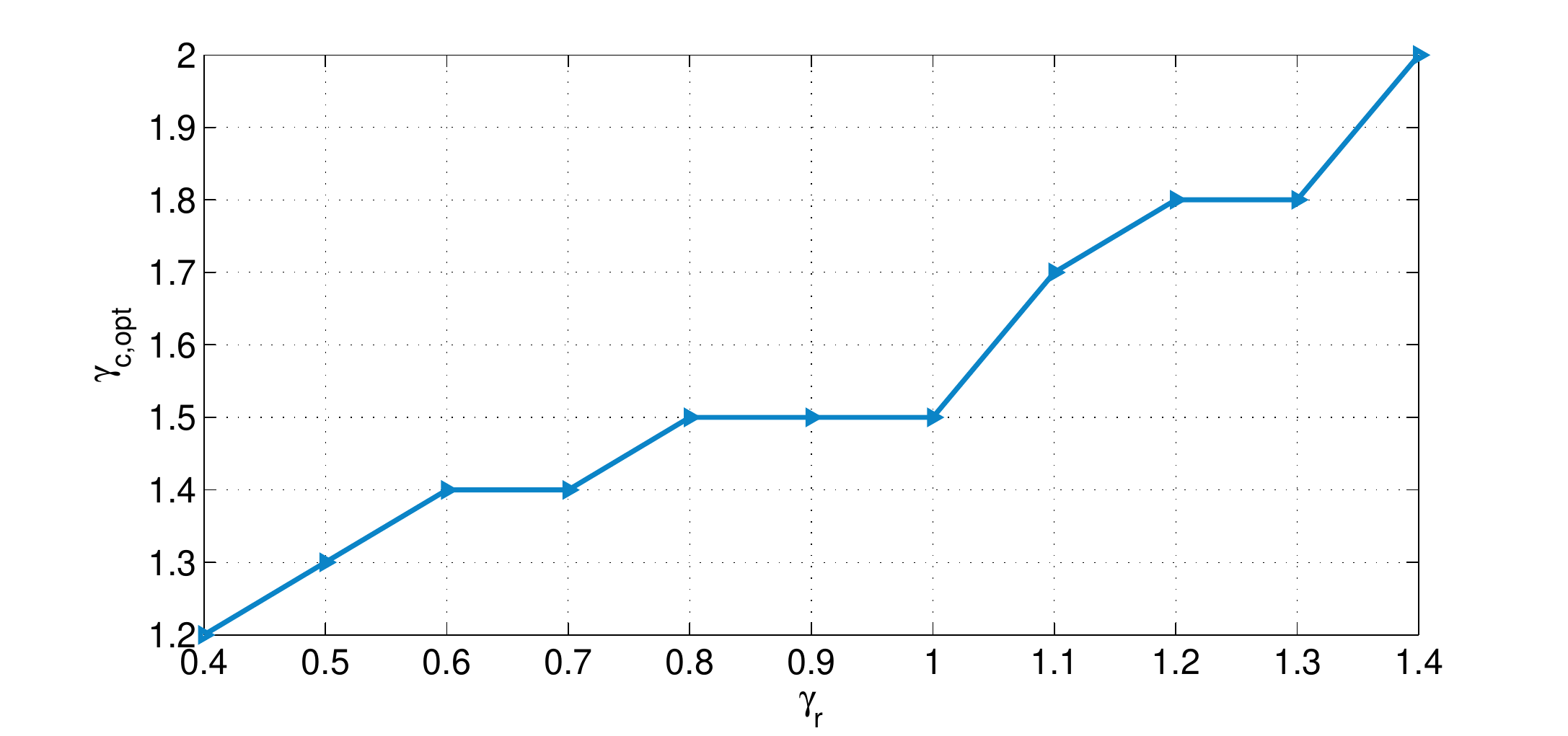}}
    \caption{The optimal $\gamma_c$ versus $\gamma_r$ for random caching with $r=r_{opt}=0.2$, $n=500$ and $m=1000$.}
    \label{fig:copt_gamma}
    \end{figure}

We compare deterministic and random caching in figures \ref{fig:random_fixed}, \ref{fig:EA_fr_m}, and \ref{fig:EA_fr_n}.
These figures confirm that deterministic caching achieves better performance than random caching. This happens because in deterministic caching the cached files do not overlap. On the other hand, random caching is based on distributed placement and there is a chance of caching the same file in different devices; as a result the performance will decrease.

  Figure \ref{fig:random_fixed} compares the average number of active clusters versus the collaboration distance for deterministic caching and random caching. For random caching, the exponent of caching distribution, $\gamma_c$ is assume to be the optimal value $1.5$.
 As it is obvious from the figure, the optimal collaboration distance for both deterministic and random caching strategies are equal but the average number of active links is higher for deterministic caching, as expected. We also compare the average number of satisfied files which is summation of the average number of D2D links and the average number of self-requests for deterministic caching, random caching, and the \emph{most-popular-caching-only} strategies.
 In the \emph{most-popular-caching-only} scheme each user stores the most popular file. Therefore, the average number of self-requests is maximized but users cannot have D2D communication because of the fact that all of them store the same file. As expected, the average number of self-requests for this simple strategy is independent of the collaboration distance. Figure \ref{fig:random_fixed} also shows that for large value of collaboration distance, $E(A+n_{self})$ is larger in the randomized caching.
The reason is that due to randomness, the chance of self-requesting is higher in random caching than in deterministic caching. Therefore, the average number of self-requests is larger than that for deterministic caching.

    \begin{figure}[htb]
    \centerline{\includegraphics[width=14cm]{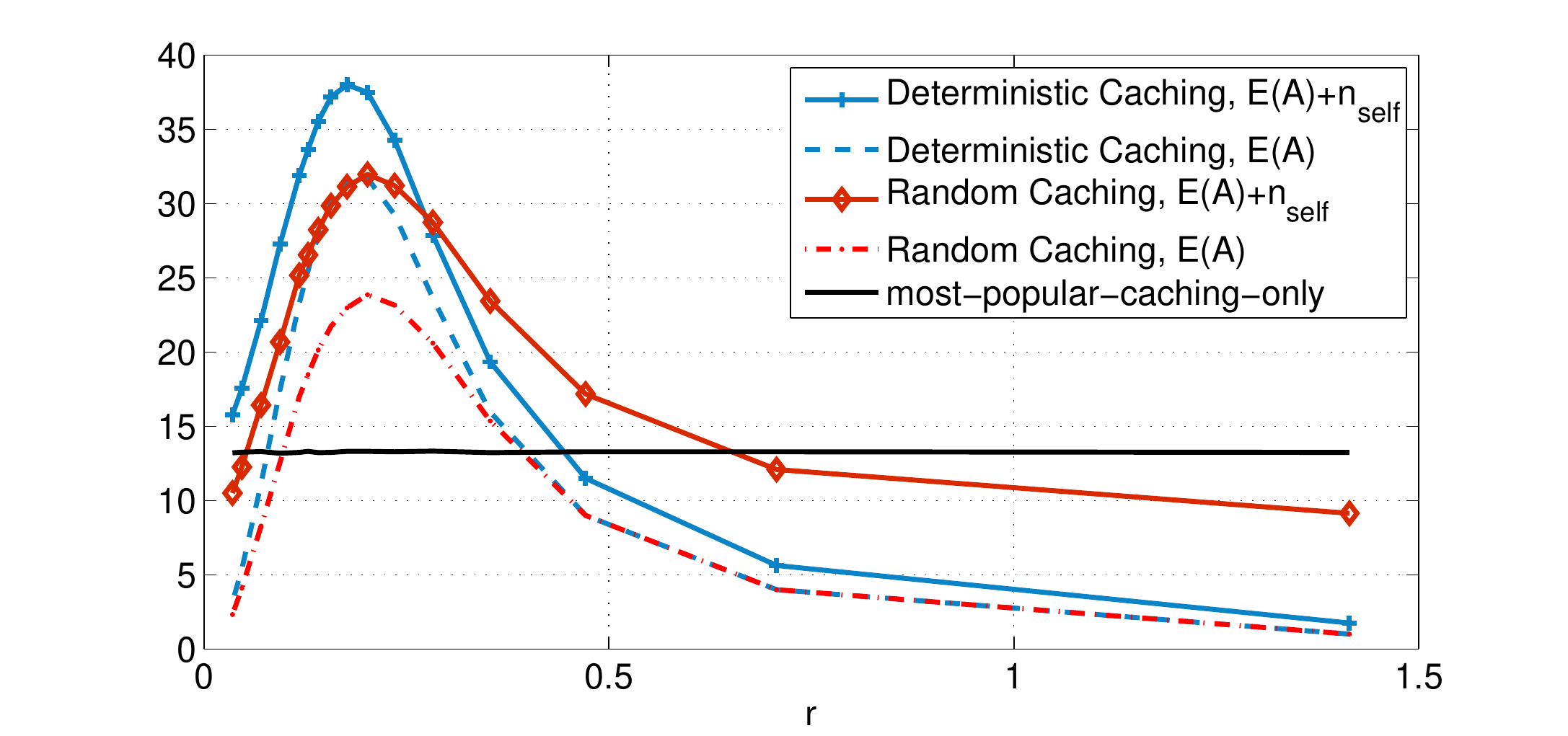}}
    \caption{The average number of active clusters and number of self-requests versus the collaboration distance for deterministic and random caching (with optimal $\gamma_c$) and one simple strategy where all users cache the most popular file, with $\gamma_r=0.6$, $n=500$ and $m=1000$.}
    \label{fig:random_fixed}
    \end{figure}

Fig. \ref{fig:nself_random_fixed} shows the average number of self-requests versus the collaboration distance for different values of $\gamma_r$. This figure confirms that as the collaboration distance increase, the ratio of self-requests of the random caching to that of the deterministic caching grows and this is more significant for larger $\gamma_r$s.

    \begin{figure}[htb]
    \centerline{\includegraphics[width=14cm]{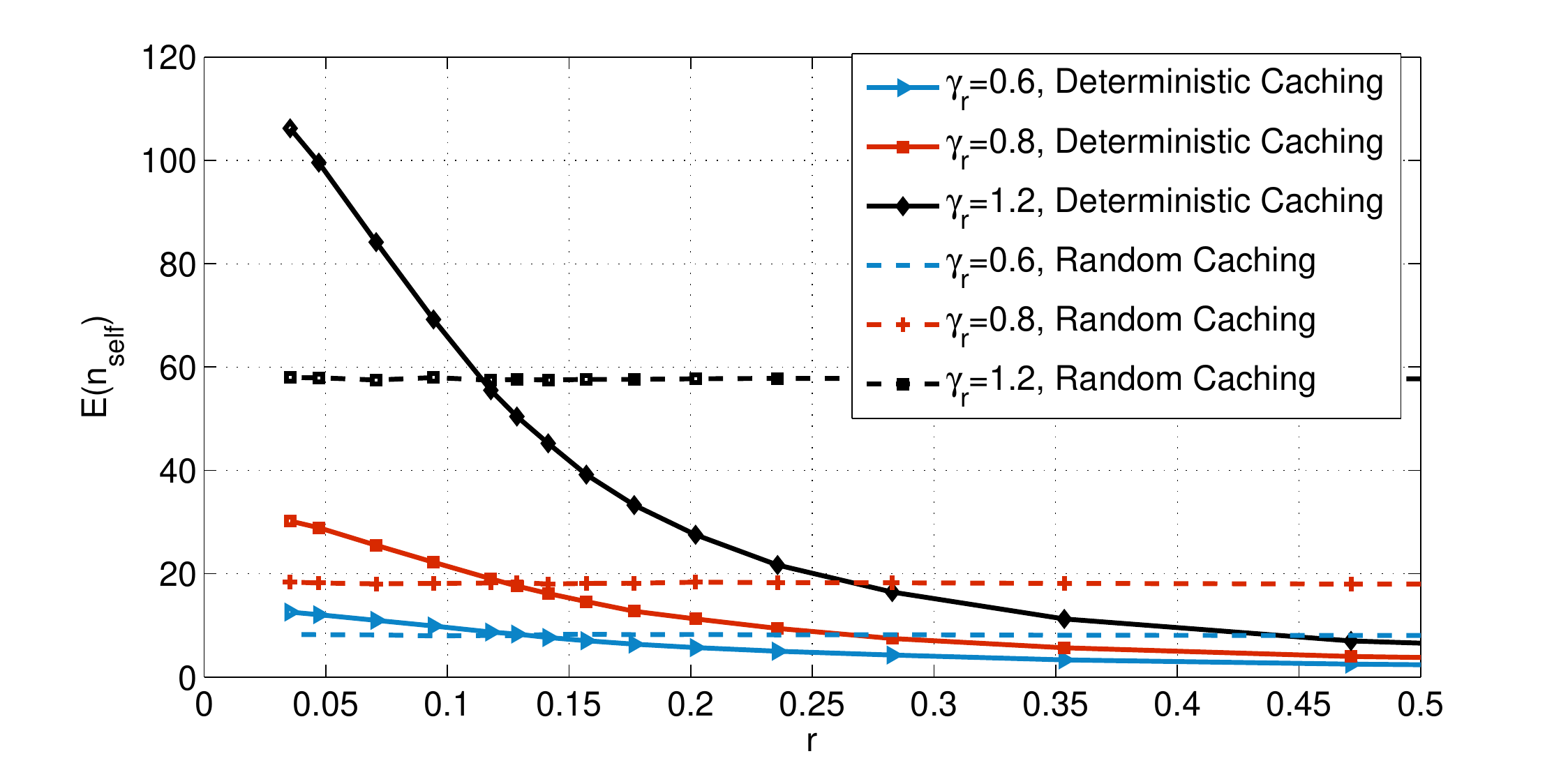}}
    \caption{The average number of self-requests versus the collaboration distance for deterministic and random caching (with optimal $\gamma_c$), for $n=500$ and $m=1000$.}
    \label{fig:nself_random_fixed}
    \end{figure}

 Figures \ref{fig:EA_fr_m} and \ref{fig:EA_fr_n} show respectively the effects of the size of the library and the number of users on the optimal average number of active links for both deterministic and random caching.
     The same arguments about the effects of $n$ and $m$ on the average number of active clusters as  discussed in figures \ref{fig:maxEA_m} and \ref{fig:maxEA_n_logx} are valid here.
   Figure \ref{fig:EA_fr_n} shows that for small number of users deterministic and random caching have very similar performance, but as n grows, the difference between their performance will \emph{increase}. The reason is that when number of users grows, the probability of having overlap between cached files in random caching strategy will increase, therefore the improvement of its performance will be less than deterministic caching.


\begin{figure}[htb]
    \centerline{\includegraphics[width=14cm]{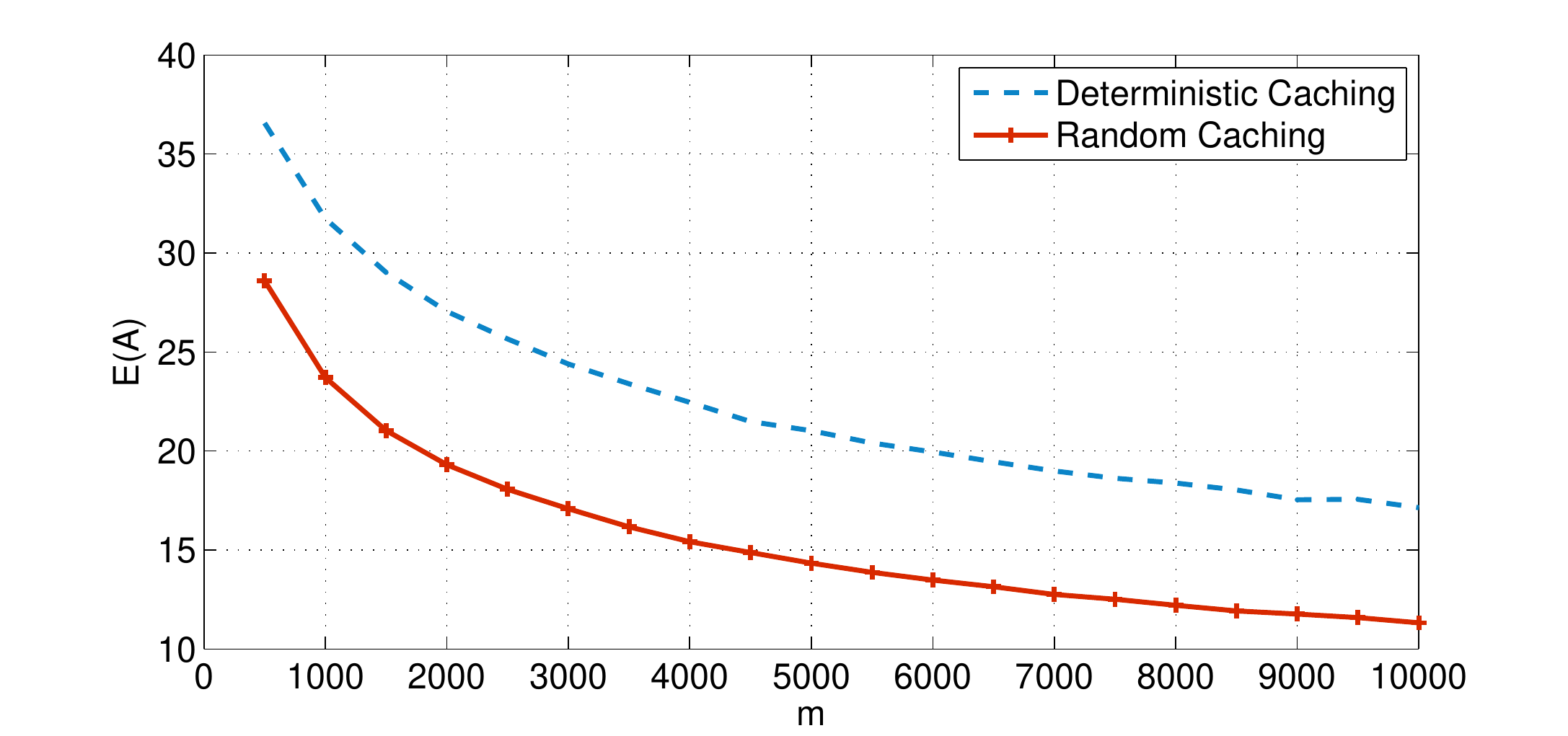}}
    \caption{The average number of clusters versus the size of the library $m$ for deterministic and random caching with $\gamma_r=0.6$, $r=0.2$ and $n=500$.}
    \label{fig:EA_fr_m}
    \end{figure}

 \begin{figure}[htb]
    \centerline{\includegraphics[width=14cm]{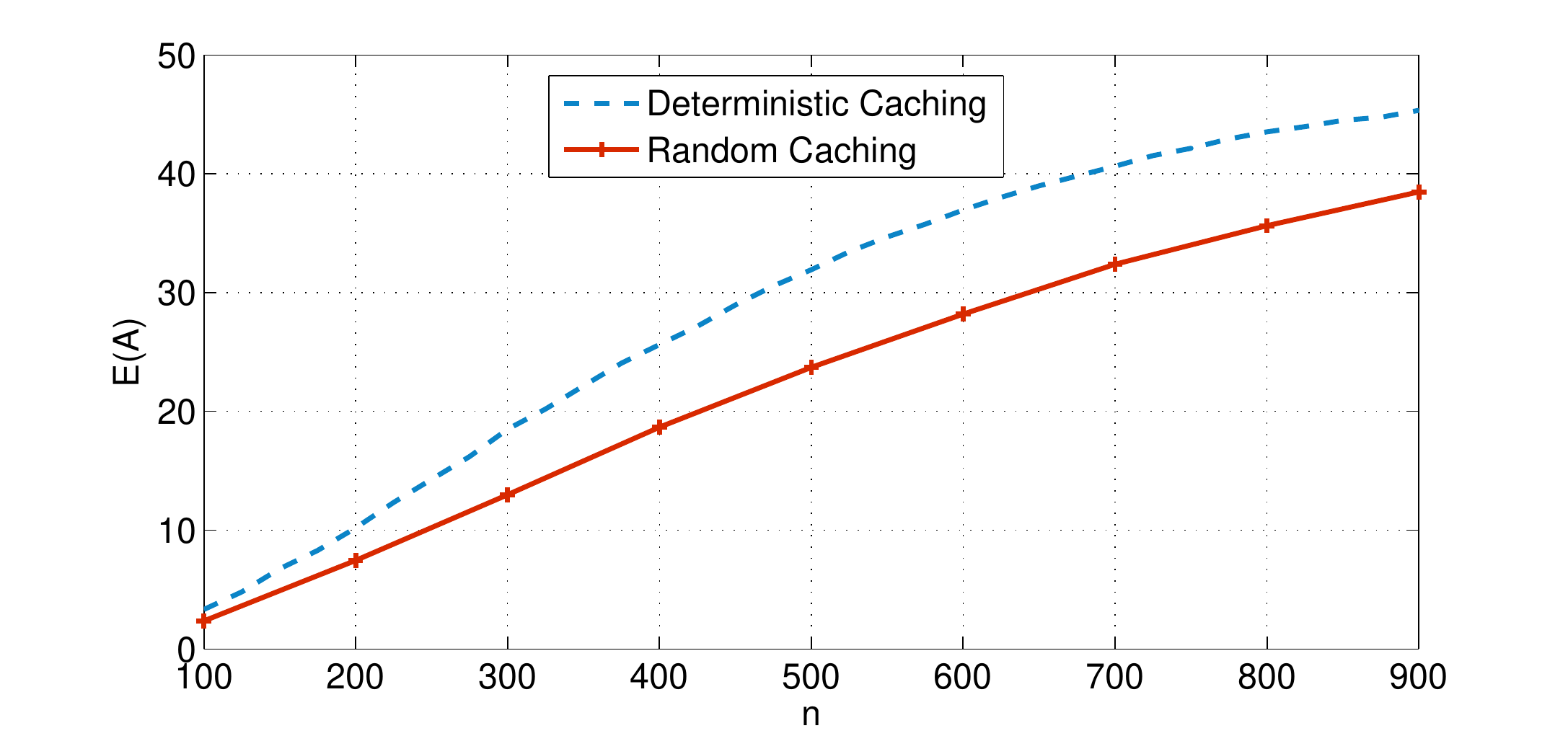}}
    \caption{The average number of clusters versus the number of users $n$ for deterministic and random caching with $\gamma_r=0.6$, $r=0.2$ and $m=1000$.}
    \label{fig:EA_fr_n}
    \end{figure}

\section{Summary and conclusions}\label{sec_conc}

The use of D2D communications has the capability of increasing the throughput of wireless video networks by orders of magnitude. When the transmission of the most popular files can be offloaded to D2D, which has high frequency reuse and thus high area spectral efficiency, the BS is freed up for providing rarely requested video files as well as non-video data. We analyzed both deterministic and random caching strategies, and pointed out that the latter is more realistic in networks with user mobility.

While this paper presented a (simplified) system model and extensive simulations, much additional work remains to be done. The question of the optimal caching distribution, as well as strategies to learn the request distribution and bring the files into the cache, are of great practical importance. Furthermore, the subdivision of cells into clusters (with no communication across cluster boundaries) might not be optimal, and more general scheduling strategies are required in that case.

Yet, our simulations have shown that even somewhat sub optimal clustering strategy can lead to orders of magnitude improvement of throughput for high user density in a cell. Our new concept thus presents a highly promising strategy for alleviating a key bottleneck in cellular data transmission.

\section{Acknowledgement}

Part of this work was supported by the Intel VAWN (Video-Award Wireless Networks) Program. We thank Dr. Chris Ramming, Dr. David Ott, and Dr. Jeff Foerster for their support, encouragement, and helpful discussions. We also thank Prof. Giuseppe Caire, Prof. Mike Neely, Prof. Antonio Ortega, and Prof. Jay Kuo, for many insightful discussions.

\bibliographystyle{IEEEtran}
\bibliography{ref_TWC}

\begin{thebibliography}{10}
\providecommand{\url}[1]{#1}
\csname url@samestyle\endcsname
\providecommand{\newblock}{\relax}
\providecommand{\bibinfo}[2]{#2}
\providecommand{\BIBentrySTDinterwordspacing}{\spaceskip=0pt\relax}
\providecommand{\BIBentryALTinterwordstretchfactor}{4}
\providecommand{\BIBentryALTinterwordspacing}{\spaceskip=\fontdimen2\font plus
\BIBentryALTinterwordstretchfactor\fontdimen3\font minus
  \fontdimen4\font\relax}
\providecommand{\BIBforeignlanguage}[2]{{%
\expandafter\ifx\csname l@#1\endcsname\relax
\typeout{** WARNING: IEEEtran.bst: No hyphenation pattern has been}%
\typeout{** loaded for the language `#1'. Using the pattern for}%
\typeout{** the default language instead.}%
\else
\language=\csname l@#1\endcsname
\fi
#2}}
\providecommand{\BIBdecl}{\relax}
\BIBdecl

\bibitem{cisco66}
``{http://www.cisco.com/en/US/solutions/collateral/ns341/ns525/ns537
  /ns705/ns827/white/paper/c11-520862.html.}''

\bibitem{surveyfemto}
V.~Chandrasekhar, J.~Andrews, and A.~Gatherer, ``Femtocell networks: a
  survey,'' \emph{Communications Magazine, IEEE}, vol.~46, no.~9, pp. 59--67,
  2008.

\bibitem{ahlehagh2012hierarchical}
H.~Ahlehagh and S.~Dey, ``Hierarchical video caching in wireless cloud:
  Approaches and algorithms,'' in \emph{Communications (ICC), 2012 IEEE
  International Conference on}.\hskip 1em plus 0.5em minus 0.4em\relax IEEE,
  2012, pp. 7082--7087.

\bibitem{maddah2013decentralized}
M.~A. Maddah-Ali and U.~Niesen, ``Decentralized caching attains order-optimal
  memory-rate tradeoff,'' \emph{arXiv preprint arXiv:1301.5848}, 2013.

\bibitem{chellouche2012home}
S.~A. Chellouche, D.~N{\'e}gru, Y.~Chen, and M.~Sidibe, ``Home-box-assisted
  content delivery network for internet video-on-demand services,'' in
  \emph{Computers and Communications (ISCC), 2012 IEEE Symposium on}.\hskip 1em
  plus 0.5em minus 0.4em\relax IEEE, 2012, pp. 000\,544--000\,550.

\bibitem{femtocaching}
N.~Golrezaei, K.~Shanmugam, A.~Dimakis, A.~Molisch, and G.~Caire,
  ``Femtocaching: Wireless video content delivery through distributed caching
  helpers,'' in \emph{INFOCOM}.\hskip 1em plus 0.5em minus 0.4em\relax IEEE,
  2012.

\bibitem{golrezaei2012wireless}
N.~Golrezaei, K.~Shanmugam, A.~G. Dimakis, A.~F. Molisch, and G.~Caire,
  ``Wireless video content delivery through coded distributed caching,''
  \emph{Communications (ICC), 2012 IEEE International Conference on}, pp.
  2467--2472, 2012.

\bibitem{wifi}
\BIBentryALTinterwordspacing
``Wifi direct demonstration at ces 2011.'' [Online]. Available:
  \url{www.youtube.com/watch?v=_mv-XFZmwNA}
\BIBentrySTDinterwordspacing

\bibitem{wu2010flashlinq}
X.~Wu, S.~Tavildar, S.~Shakkottai, T.~Richardson, J.~Li, R.~Laroia, and
  A.~Jovicic, ``Flashlinq: A synchronous distributed scheduler for peer-to-peer
  ad hoc networks,'' in \emph{Communication, Control, and Computing (Allerton),
  2010 48th Annual Allerton Conference on}.\hskip 1em plus 0.5em minus
  0.4em\relax IEEE, 2010, pp. 514--521.

\bibitem{doppler2010mode}
K.~Doppler, C.-H. Yu, C.~B. Ribeiro, and P.~Janis, ``Mode selection for
  device-to-device communication underlaying an lte-advanced network,'' in
  \emph{Wireless Communications and Networking Conference (WCNC), 2010
  IEEE}.\hskip 1em plus 0.5em minus 0.4em\relax IEEE, 2010, pp. 1--6.

\bibitem{gupta2000capacity}
P.~Gupta and P.~Kumar, ``The capacity of wireless networks,'' \emph{Information
  Theory, IEEE Transactions on}, vol.~46, no.~2, pp. 388--404, 2000.

\bibitem{keller2012microcast}
L.~Keller, A.~Le, B.~Cici, H.~Seferoglu, C.~Fragouli, and A.~Markopoulou,
  ``Microcast: Cooperative video streaming on smartphones,'' in
  \emph{Proceedings of the 10th international conference on Mobile systems,
  applications, and services}.\hskip 1em plus 0.5em minus 0.4em\relax ACM,
  2012, pp. 57--70.

\bibitem{VAWN_2010}
``{http://software.intel.com/en-us/articles/video-aware-wireless-networks.}''

\bibitem{golrezaei2012base}
N.~Golrezaei, A.~F. Molisch, and A.~G. Dimakis, ``Base-station assisted
  device-to-device communications for high-throughput wireless video
  networks,'' in \emph{Communications (ICC), 2012 IEEE International Conference
  on}.\hskip 1em plus 0.5em minus 0.4em\relax IEEE, 2012, pp. 7077--7081.

\bibitem{tracedata}
``http://traces.cs.umass.edu/index.php/network/network.''

\bibitem{RGGbook}
M.~Penrose and O.~U. Press, \emph{Random geometric graphs}.\hskip 1em plus
  0.5em minus 0.4em\relax Oxford University Press Oxford, 2003, vol.~5.

\bibitem{zipf}
M.~Cha, H.~Kwak, P.~Rodriguez, Y.~Ahn, and S.~Moon, ``I tube, you tube,
  everybody tubes: analyzing the world's largest user generated content video
  system,'' in \emph{Proceedings of the 7th ACM SIGCOMM conference on Internet
  measurement}.\hskip 1em plus 0.5em minus 0.4em\relax ACM, 2007, pp. 1--14.

\bibitem{IT_D2D_paper}
N.~Golrezaei, A.~Dimakis, and A.~Molisch, ``Wireless device-to-device
  communications with distributed caching,'' \emph{ISIT 2012}.

\bibitem{ji2013optimal}
M.~Ji, G.~Caire, and A.~F. Molisch, ``Optimal throughput-outage trade-off in
  wireless one-hop caching networks,'' \emph{arXiv preprint arXiv:1302.2168},
  2013.

\bibitem{Molisch_book_2011}
A.~F. Molisch, \emph{Wireless communications}.\hskip 1em plus 0.5em minus
  0.4em\relax 2nd ed., IEEE-Press Wiley, 2011.

\bibitem{golrezaei2012femtocaching}
N.~Golrezaei, A.~F. Molisch, A.~G. Dimakis, and G.~Caire, ``Femtocaching and
  device-to-device collaboration: A new architecture for wireless video
  distribution,'' \emph{arXiv preprint arXiv:1204.1595}, 2012.

\end{thebibliography}

\end{document}